\documentclass[aps,prl,twocolumn,superscriptaddress]{revtex4-2}

\usepackage{amssymb}
\usepackage{amsmath}
\usepackage{graphicx}
\usepackage{xspace}		
\usepackage[separate-uncertainty=true]{siunitx}
\usepackage{xcolor}
\usepackage[nolist]{acronym}
\usepackage[]{overpic}
\usepackage[normalem]{ulem}  
\usepackage{color,soul}  
\usepackage{tikz}
\usetikzlibrary{calc}
\usepackage{easyReview}

\newcommand{\zpinch}{Z-pinch\xspace}
\newcommand{\zpinches}{Z-pinches\xspace}
\newcommand{\btheta}{$B_{\theta}$\xspace}

\newacro{LDP}{low-density plasma}
\newacro{SP}{stagnating plasma}
\newacro{ICCD}{intensified charged coupled device}
\newacro{WIS}{Weizmann institute of Science, Israel}
\newacro{IAT}{inverse Abel transform}
\newacro{CR}{collisional radiative}
\newacro{CRE}{collisional radiative equilibrium}
\newacro{MHD}{magnetohydrodynamics}
\newacro{MRT}{magneto-Rayleigh-Taylor}
\newacro{LoS}{line of sight}
\newacro{LCE}{lumped-circuit equation}

\begin{document}



\title{	Observation of fast current redistribution in an imploding plasma column }


\author{C.~Stollberg}
\email[Corresponding author: ]{christine.stollberg@epfl.ch}
\author{E.~Kroupp}
\author{D.~Mikitchuk}
\author{P.~Sharma}
\author{V.~Bernshtam}
\author{M.~Cveji\'{c}}
\author{R.~Doron}
\author{E.~Stambulchik}
\author{Y.~Maron}

\affiliation{Weizmann Institute of Science, Herzl street 243, 7610001 Rehovot, Israel}

\author{A.~Fruchtman}
\affiliation{Department of Physics, Holon Institute of Technology, Holon 58102, Israel}

\author{I.~E.~Ochs}
\author{N.~J.~Fisch}
\affiliation{Department of Astrophysical Sciences, Princeton University, Princeton, New Jersey 08540, USA}

\author{U.~Shumlak}
\affiliation{Aerospace \& Energetics Research Program, University of Washington, Seattle, Washington 98195, USA}

\date{\today}

\begin{abstract}

Spectroscopic measurements of the  magnetic field evolution in a \zpinch throughout stagnation and with particularly high spatial resolution reveal a sudden current redistribution from the stagnating plasma (SP) to a low density plasma (LDP) at larger radii, while the SP continues  to implode. Based on the plasma parameters  it is shown that the current is transferred to an increasing-conductance LDP outside the stagnation, a process likely to be induced by the large impedance of the SP. Since an LDP often exists around imploding plasmas and in various pulsed-power systems, such a fast current redistribution may dramatically affect the behavior and achievable parameters in these systems.

\end{abstract}



\maketitle

\textit{Introduction - } 
The force exerted on a plasma by a current and its associated magnetic field governs various processes in space and laboratory plasmas \cite{burt2004lightning, chen2017sun, tzeferacos2018dynamo, kikuchi2010fusionReview, sinars2020review, ryutov2000review, giuliani2015review, slutz2012prl, gomez2014maglif, maron2020tutorial, black2000pos, engelbrecht2018pos,  weber1995pos, gomez2017currentLoss, bennett2021transmissionLines, lebedev2019astrophysics}.  The implosion of a current-carrying plasma column, as in \zpinches, is a typical example where the current distribution in the plasma plays a crucial role \cite{giuliani2015review, sinars2020review, ryutov2000review}. In this configuration, the plasma is generated by a pulsed current driven axially in a cylindrical gas or a metallic load. The current generates an azimuthal magnetic field \btheta that accelerates the plasma radially inward and implodes together with the plasma \cite {giuliani2015review, sinars2020review, ryutov2000review, slutz2012prl, gomez2014maglif, maron2020tutorial}. Thus, a \zpinch is a platform for investigating the interplay between the current density distribution and the evolution of the implosion and stagnation, that are highly complex \cite{sinars2020review, ryutov2000review, giuliani2015review, maron2020tutorial, alumot2019ionTemperature, kroupp2018turbulent, lebedev2004imperialCollege}.

A detailed analysis of the energy and pressure balance of 
two \zpinch plasmas, distinctly different in power \cite{maron2013}, revealed that, in both the low- and high-power systems, the magnetic pressure due to $B_{\theta }$ is small at the \ac{SP}.  Contrary to what had been believed, at most 1/3 of the discharge current was found to flow within the  \ac{SP}.
The direct confirmation of this result required experimental measurements of \btheta during the stagnation and close to the \ac{SP}, a task that is challenging due to
the high electron density $n_{e}$, the high ion velocities, and the
transient nature of the plasma \cite{doron2014bfield}.

A few studies reported on the direct measurement of magnetic fields in \zpinches \cite{davara1998, gomez2014zeeman, mikitchuk2019, aybar2022, rosenzweig2017jinst, rosenzweig2020pop, ivanov2015faraday, munzar2021deflectometry}.
Those performed at large radii during the early stage of implosion \cite{davara1998, gomez2014zeeman, mikitchuk2019, aybar2022} demonstrate that in general the discharge current flows in the imploding plasma. Also, studies performed at the time of stagnation \cite{rosenzweig2020pop, rosenzweig2017jinst, ivanov2015faraday}, outside  the SP but close enough to it, confirmed that only a small fraction of the current flows in the compressed \ac{SP}, while the majority of the current flows  at larger radii.
In the present work, we address the mechanism that leads to the low current at stagnation by measuring the temporal evolution of the current density radial distribution throughout the entire implosion and stagnation, down to the small radius of the \ac{SP}, in particular with a high spatial resolution. 

The measurements  revealed that in the section close to the cathode: 
  1) the entire current flows in the imploding plasma until stagnation, 
  2) the current then suddenly redistributes from the small radius of the stagnation to larger radii where a hollow \ac{LDP} resides, and 
  3) while the current transitions to larger radii, the \ac{SP} continues to implode.

This detailed  observation of the fast current redistribution was obtained from temporally and radially  resolved measurements  of the local magnetic field  in a  small-scale gas-puff \zpinch, where the initial gas distribution was peaked on axis, thus allowing for satisfactory reproducibility and symmetry by mitigating the Magneto-Rayleigh-Taylor instabilities \cite{chang1991gasDistribution, velikovich1998tailored}.
The diagnostics, based on Zeeman-polarization spectroscopy, allowed for
determining $B_{\theta }$, even when the Zeeman-split pattern was 
obscured by the line broadening \cite{davara1998, golingo2010NoteZS, rosenzweig2020pop, rosenzweig2017jinst, mikitchuk2019}. 
Also, the radial distribution of charge states in the plasma was used as in  \cite{davara1998, rosenzweig2020pop, rosenzweig2017jinst}  to obtain $B_{\theta}$ locally as a function of $r$, from chordal measurements, without the need to inverse Abel transform the observed line shapes (no azimuthal uniformity is assumed) \cite{rosenzweig2017jinst, rosenzweig2020pop}. 
Also, the measurements of the plasma parameters  allowed for inferring the plasma resistivity required for the interpretation of the present results. 

Our findings are likely related to the previous observations of  low current in the \ac{SP} of \zpinches \cite{maron2013, maron2020tutorial, rosenzweig2017jinst, rosenzweig2020pop, ivanov2015faraday} 
and the phenomenon of 'trailing' current  \cite{lebedev2002snowplow, lebedev2005, waisman2004inductance, cuneo2005,  hall2006,  burdiak2013current,  rosenzweig2017jinst, rosenzweig2020pop, mikitchuk2019, maron2020tutorial, aybar2022},
and  might also be relevant for lower-density plasma systems \cite{maron2020tutorial},  such as plasma switches \cite{black2000pos, engelbrecht2018pos, weber1995pos} and high power transmission lines \cite{gomez2017currentLoss, bennett2021transmissionLines}.

\textit{Experimental Methods - } In the present experiment  
\cite{stollberg2019phd}, an oxygen column of initial diameter $\approx$ \SI{3}{mm}  and  mass of \SI{1.6 \pm 0.6}{\micro\gram/\centi\meter} is injected into a 6-mm-long anode-cathode gap and imploded by a current pulse rising to \SI{27}{kA} in \SI{160}{ns}.

\begin{figure*}
\includegraphics[width=\textwidth]{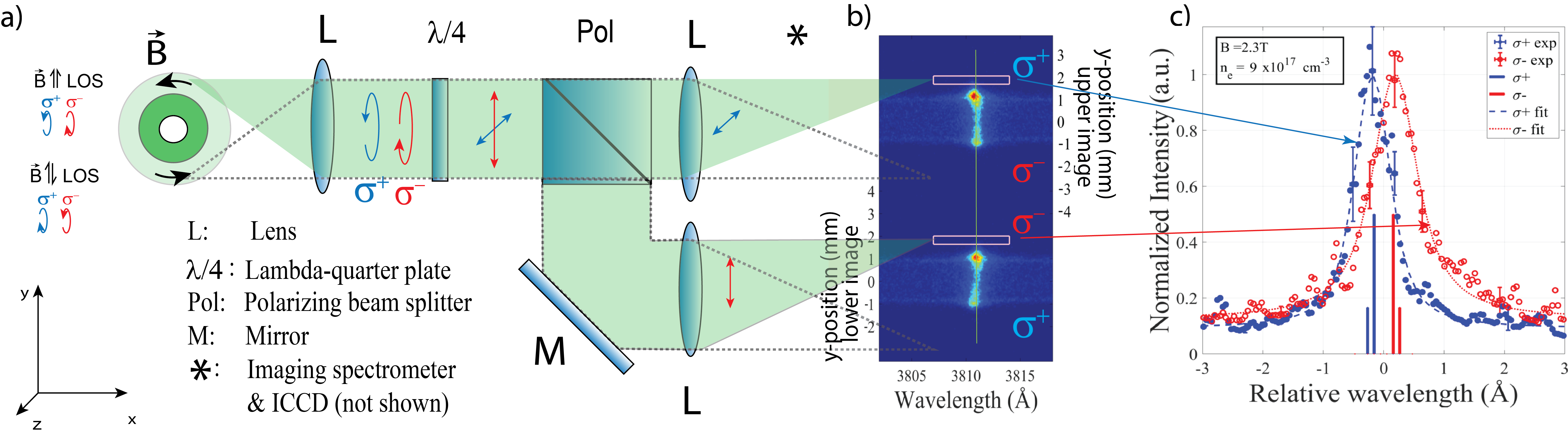}
	\caption{\label{fig:diagnostics}
	Schematics for B-field determination:
		a) Diagnostic setup: With the side-on imaging, at the outermost chord, the \ac{LoS} is parallel to $\vec{B}$ so that only the circularly polarized $\sigma^+$ and  $\sigma^-$ Zeeman components are observed. The light is transformed into orthogonal, linearly polarized light by a $\lambda$/4 plate and spatially separated by a polarizing beam splitter. The two plasma images, with the different polarization properties, are defracted by a 1-m imaging spectrometer and recorded on an \ac{ICCD} (not shown on the image). 
		b) Typical spectral recording of the O VI 3s - 3p transition obtained at t = \SI{100}{ns}. The green line represents the unshifted spectral line, illustrating the deviation of the measured spectral lines from the unperturbed position, where $\sigma^+$  and $\sigma^-$  exhibit blue and red (B-dependent) shifts, respectively. 
		c) Spectral lineouts (markers) taken from b) at y = \SI{1.6 \pm 0.1}{mm} from the upper and the lower image (bright rectangles) and their fit (dashed/dotted). The solid lines represent the calculated Zeeman pattern without line broadening. The separation between the profiles of \SI{0.37}{\angstrom} yields a  B-field of \SI{2.3}{T}, and the Stark broadening yields $n_e \approx$  \SI{9e17}{\per\centi\meter\cubed} \cite{stambulchik2006broadening, stambulchik2008stark}.}
		
\end{figure*}

For determining the B-field, the wavelength separation between the $\sigma^+$ and the $\sigma^-$ Zeeman components is measured \cite{doron2014bfield}.
Employing a side-on imaging (Fig. \ref{fig:diagnostics}) the \ac{LoS} through the outermost chord of a spectral line emission is parallel to \btheta, which allows for spatially separating the $\sigma$ Zeeman components and recording them simultaneously on a single detector coupled to an imaging spectrometer. 
The spectroscopic system  spectral, temporal, and spatial resolutions are \SI{0.3}{\angstrom}, \SI{10}{ns}, and \SI{100}{\mu m} ($\approx$ 1/5 of the SP radius), respectively.
In order to obtain \btheta at different radii, we use up to three different spectral lines that are radially separated \cite{rosenzweig2017jinst}. 

A second imaging spectroscopic setup, providing a broader spectrum \cite{stollberg2019phd}, is used for temperature (from  line intensity ratios \cite{gregorian2005temperature, gregorian2005ionization, ralchenko2001nomad}) and density (from Stark broadening \cite{stambulchik2006broadening, stambulchik2008stark, konjevic2002stark} and absolute intensity \cite{gregorian2005ionization}) diagnostics, as well as the estimation of the average charge state $Z_\mathit{eff}$, which yields the plasma resistivity.

\begin{figure}
	\includegraphics[width=0.5\textwidth]{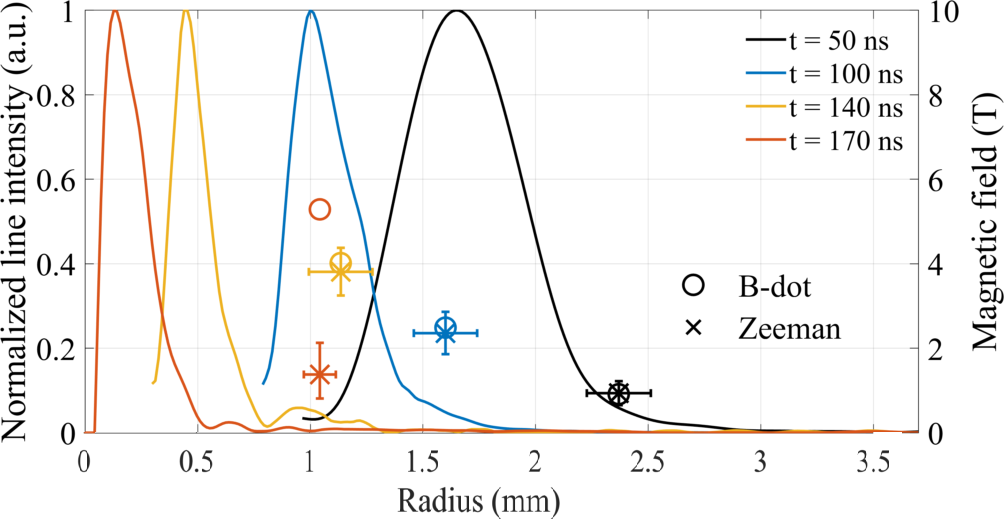}
	\caption{\label{fig:B_outer_radius} Evolution of the O~IV (t = 50 ns) and O~VI (t $\geq$ \SI{100}{ns}) emission intensity radial distribution along with the magnetic field measured at the outer chord: at t = \SIrange{50}{140}{ns}, the agreement between expected  (B-dot) and  measured magnetic field (Zeeman) indicates the confinement of the entire discharge current in the imploding plasma. At t = 170 ns the fraction of the current that can be located in the stagnating plasma is $\lesssim$ 30\%.}
\end{figure}

\textit{Results - }  
Presented are results obtained at  z = \SI{1}{mm} from the cathode valve.  Fig. \ref{fig:diagnostics} b shows a typical spectral recording of the chord-integrated plasma emission, split into two images with different polarizations, as described above.
To determine the B-field, spectral lineouts (Fig. \ref{fig:diagnostics} c) are extracted at the outer edge of the O~VI spectral line and fitted with Voigt-profiles.
The fitting accounts for the Zeeman components (\SI{0.1}{\angstrom}), the  Gaussian contributions from the instrumental and Doppler broadenings (\SI{0.4}{\angstrom}) and the Lorentzian contribution (\SI{0.6}{\angstrom}) from  the Stark broadening. For the  3s - 3p transition of O~VI at \SI{3811.35}{\angstrom}, the separation of the Zeeman $\sigma$ components is \SI{0.16}{\angstrom}  for $B = $ \SI{1}{T} \cite{doron2014bfield}. For the example in Fig. \ref{fig:diagnostics} c we find \btheta~=~$2.3_{-0.2}^{+0.4}$ T, where the uncertainty includes the fitting error, spectrometer aberrations, and the performance of the polarization optics \cite{stollberg2019phd, rosenzweig2017jinst}. 
The Stark broadening allows for determining  the electron density  \SI{0.9 \pm 0.3 e18}{\per\centi\meter\cubed} \cite{stambulchik2006broadening, stambulchik2008stark}.
The integration region in the radial direction is defined for each spectral lineout individually, depending on signal-to-noise ratio and line broadening, and is typically \SIrange{100}{200}{\micro\meter}.

Fig. \ref{fig:B_outer_radius} summarizes the results of the B-field measurements at different times during the implosion (t $<$ \SI{140}{ns}) and at stagnation (t =\SIrange{140}{170}{ns}), demonstrating the fast current redistribution. 
For any given time, used is the charge state that resides at the outermost radius of the imploding plasma. Its emission-intensity radial distribution is shown along with the \btheta values obtained at the outer edge $r_{out}$ of its line emission.
Also shown is \btheta obtained by assuming that the entire current $I(t)$, measured by a calibrated B-dot probe (90\% of $I(t)$ are considered as the entire current), flows within the radius of the B-field measurement, namely $B_{B-dot}(t) = 0.9 \ \mu_0 \, I(t) / 2 \pi r_{out}(t)$.
We note that each data point was obtained in a separate discharge, with shot-to-shot variations  significantly smaller than the error bars quoted in Fig. \ref{fig:B_outer_radius}.
For t $\leq$ \SI{140}{ns}, the data show a good agreement between  the measured  (Zeeman) and the expected (B-dot) B-field values, indicating that the entire discharge current flows in the imploding plasma. At t = \SI{170}{ns}, however, a sig\-ni\-fi\-cant discrepancy is observed: at most, 30\% of the discharge current flows within the stagnating plasma that is located at r $<$ \SI{1.1}{mm}.

At the time of stagnation, we also use additional transitions from the \ac{LDP} to obtain the current density distribution at  $r >$ \SI{1.1}{mm}.
The O~IV 3p - 3d transition at \SI{3736.85}{\angstrom} is used to obtain a B-field of \SI{1.7}{T} at r = \SI{1.9 \pm 0.35}{mm}, corresponding to 70\% of the discharge current.
The  O~III 3s - 3p transition at \SI{3759.87}{\angstrom}, residing at $r >$ \SI{2.3}{mm}, yields a B-field of $\geq$\SI{1.3}{T} at r = \SI{3.55 \pm 0.21}{mm}, corresponding to the entire discharge current.

\begin{figure}
	\includegraphics[width=0.48\textwidth]{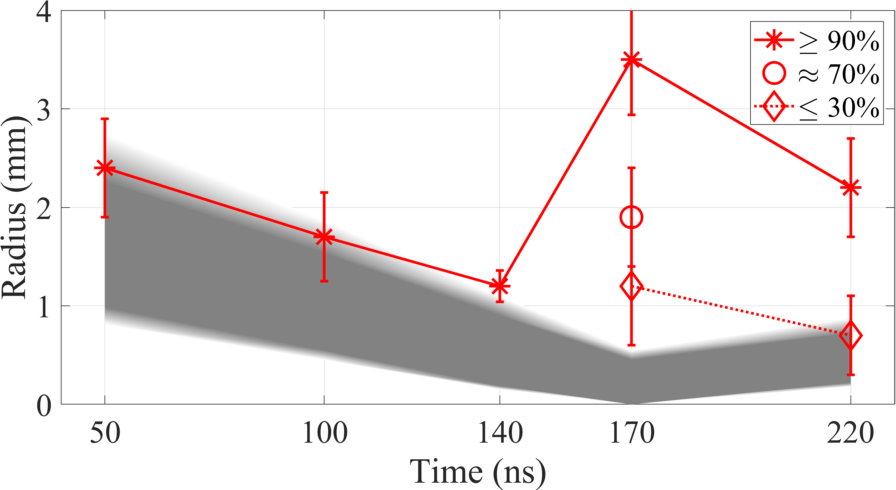}
	\caption{\label{fig:current_evolution}  Gray: location of the plasma  with $n_e > 0.1 \  n_{e, max}$ (from continuum radiation intensity) with uncertainty in light gray. Red: locations at which a certain percentage of the discharge current was measured. At t $\geq$ \SI{170}{ns} the measurements were obtained in an \ac{LDP} with a density lower than 10\% of  $n_{e,max}$. Initially, the current implodes together with the main plasma, while at stagnation (\SIrange{140}{170}{ns}), a sudden current redistribution to much larger radii takes place. }
\end{figure}

To demonstrate the correlation between the current evolution and the plasma implosion, 	
we plot in Fig. \ref{fig:current_evolution} the position of the imploding plasma (gray area) along with the positions at which the entire discharge current was measured (red line). 
Additionally, for t $\geq$ \SI{170}{ns}, we  show the positions that enclose $\approx$70\%  and $\leq$ 30\%  of the discharge current.
It is seen that during the implosion ($t <$ 140 ns), the current implodes together with the main plasma, while at stagnation (t = \SIrange{140}{170}{ns}) a current redistribution occurs to an LDP at larger radii. This demonstrates a non-monotonic evolution of the current channel, which is initially shrinking in radius, followed by a fast expansion.
Remarkably, the stagnating plasma continues its inward motion as the current  moves outwards and remains outside the SP for several tens of nanoseconds. Subsequently, the \ac{SP} expands due to thermal pressure while an inward motion of the \ac{LDP} is observed.

\textit{Discussion - }
It was shown that the entire driven current is carried by the imploding plasma until the stagnation, before a rapid transfer of the current from the SP to a LDP region at the plasma periphery occurs, revealing a transition that has not been observed  as yet.

We now turn to examine possible mechanisms for the current transfer.  
Recently, it was shown that in a zero-resistivity plasma, the current channel is expected to move radially outward when the imploding plasma velocity profile $v_r(r)$ satisfies $\partial (|v_{r}|/r)/\partial r<0$ \cite{ochs2019current}.
This condition is satisfied  up to stagnation at 170 ns, as the LDP hardly moves. To explain the rapid current redistribution that begins only after 140 ns, resistive effects must be considered. Indeed, we find that it is the rapid resistance-change in the LDP that can explain the observed fast current redistribution.

\begin{figure}
	\includegraphics[width=0.45\textwidth]{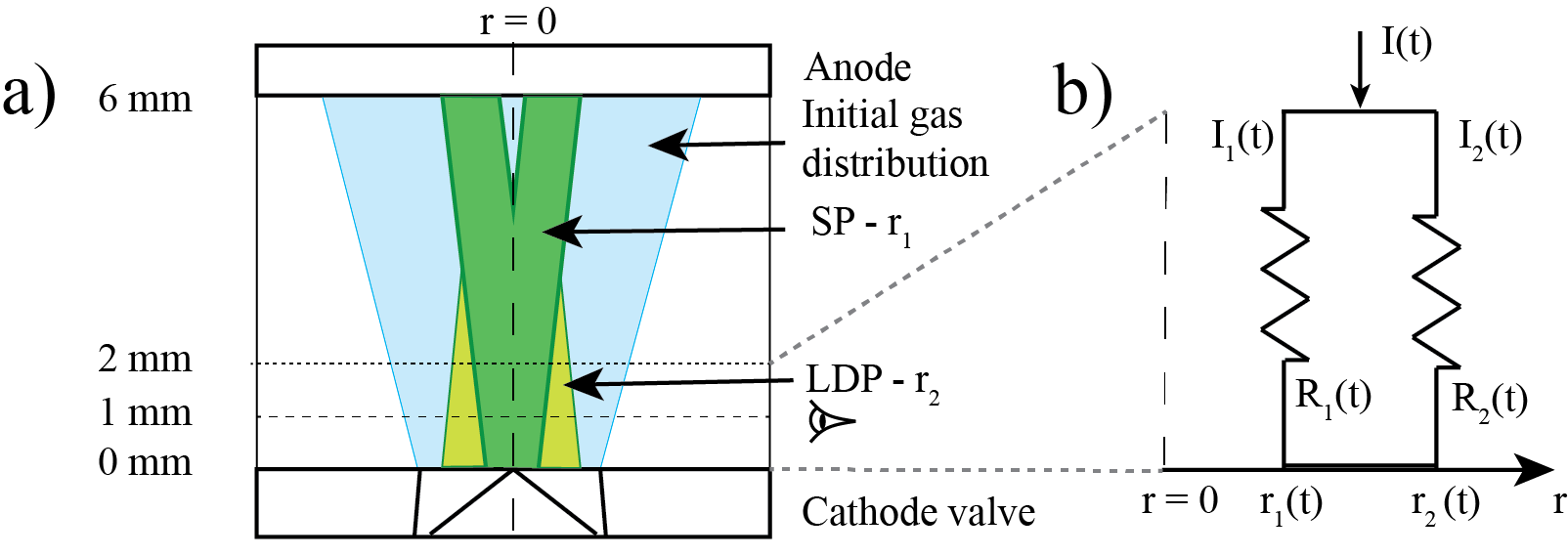}
	\caption{\label{fig:sketch}a) Schematic of the \zpinch experiment with  initial gas distribution, imploding/stagnating plasma (SP) and LDP featuring 'zippering' effect.  b) Simplification of the PE into a lumped-circuit model.}
\end{figure}

We assume that, initially, the LDP is a weakly-ionized gas (left behind from the initial gas distribution) with a resistivity much larger than that of the SP. An electric field is induced in the LDP due to the SP implosion that increases $T_e$ and leads to a fast ionization and drop of the LDP resistance below the SP impedance.

To model the process of the partitioning of the discharge current $I = I_1 + I_2$ between $I_1$, the current through the \ac{SP}, and $I_2$, the current through the \ac{LDP}, we use a \ac{LCE}:
$d(LI_{1})/dt=-I_{1}R_{1}+I_{2}R_{2}$, 
representing the circuit illustrated in Fig. \ref{fig:sketch}, where $R_{1}$ and $R_{2}$ are the resistances of the \ac{SP} and the \ac{LDP}, respectively, along $l =$ \SI{2}{mm} near the cathode.
We write the inductance as 
$L=\left( \mu _{0}l/2\pi \right) \ln \left( r_{2}/r_{1}\right) $, 
where $r_{1}$ and $r_{2}$ are the radii of the SP and of the LDP (the LDP is assumed to be relatively thin, as will be supported by experimental data elsewhere).

To illustrate the current partitioning $I_2/I$, the LCE can also be written  as $%
L \, dI_{2}/dt=-\left( Z_{1}+R_{2}\right) I_{2}+Z_{1}I$, where $Z_{1}=R_{1}+dL/dt$  is the SP impedance.
Thus, the LCE captures the physics of both, the inductive current escape due to dL/dt $>$ 0 during implosion \cite{ochs2019current},  and the resistive model \cite{haines1959skinEffect}.
The \ac{LDP} resistance is $%
R_{2}=\left( m_e/e^{2}\right) \left( l/A\right) \left[ \left( k_{eN}+\beta
\right) N/n_e+k_{ei}\right] $, where $A$ is the LDP cross section, $k_{eN}(T_e)$ and $k_{ei}(T_e)$ are the temperature dependent electron-neutral and electron-ion collision rate coefficients  and $\beta(T_e)$ is the ionization rate coefficient \cite{lieberman2005principles}, $N$ is the
density of atoms in the LDP, and $m_e$ is the electron mass.

Using the measured $r_{1}(t)$ (Fig. \ref{fig:current_evolution}),   the approximately constant $r_2 = $ \SI{2}{mm} and $I =$ \SI{20}{kA} (for \SI{100}{ns} $< t <$ \SI{220}{ns}) observed, and $R_1(t)$ as inferred from the measured SP parameters \cite{stollberg2019phd}, we solve the equations for $I_{2}$ together with $dn_e/dt=\beta Nn_e$.
To determine $R_2(t)$, a li\-ne\-ar rise-in-time of $T_e$ from   \SI{0.5}{eV} at t = \SI{100}{ns} to \SI{12}{eV} at t = \SI{220}{ns} was assumed for the LDP, consistent with the measured $T_e =$ \SI{10 \pm 2}{eV} at $t = $\SI{170}{ns} \cite{stollberg2019phd}.

\begin{figure}
	\includegraphics[width=0.5\textwidth]{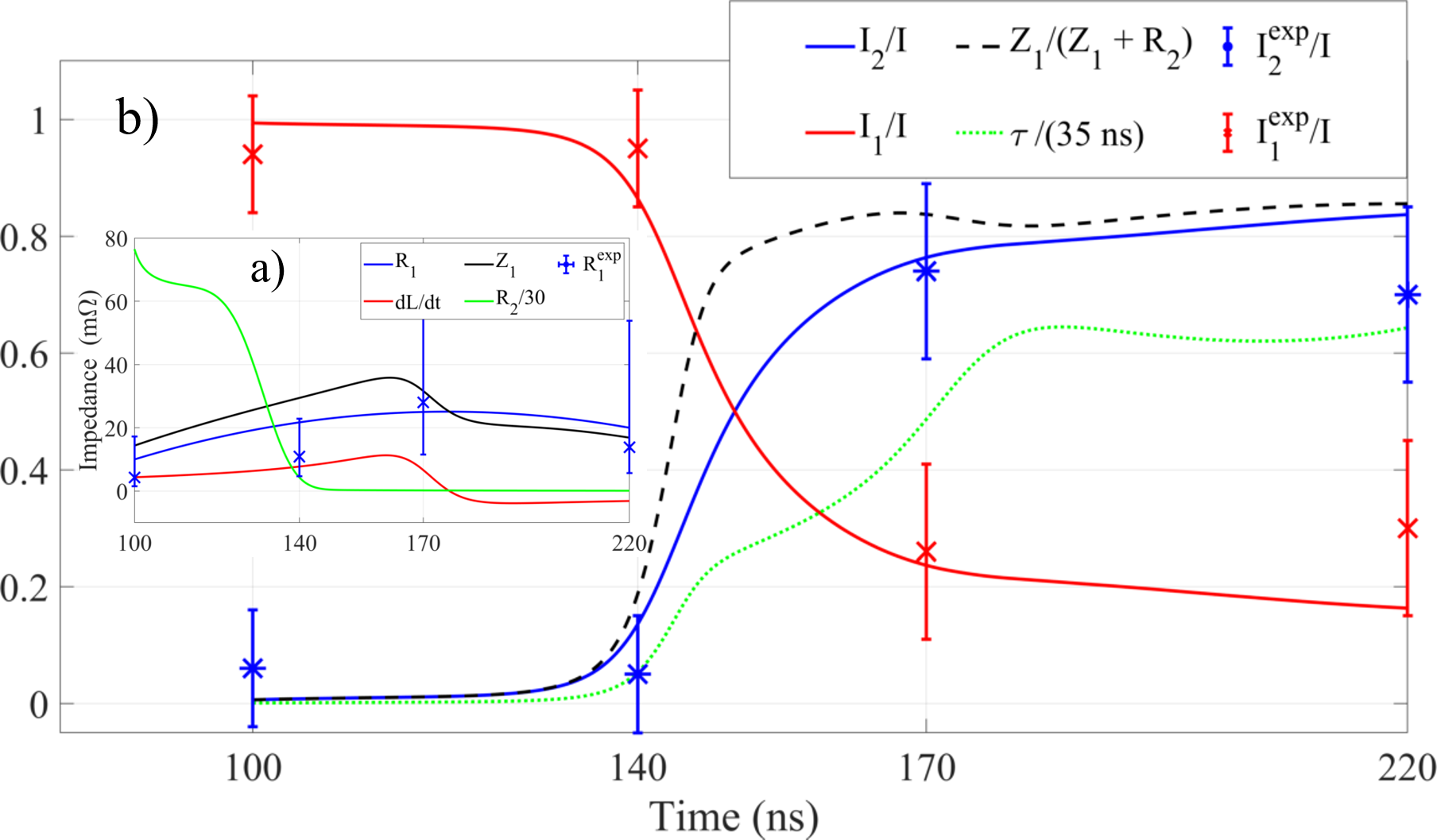}
	\caption{\label{fig:model}  Simulation of the \acf{LCE} a) the abrupt decrease of the LDP resistance $R_2$ (green) enables the current transfer, b) the evolution of the current partition $I_2/I$ (blue) follows the impedance ratio (dashed black) with a delay of the current transfer time (dotted green).} 
\end{figure}

The simulated current evolution between \SI{100}{ns} - \SI{220}{ns}, shown in Fig. \ref{fig:model}, occurs 	on three time-scales: the SP implosion time, the ionization time of the LDP, and the current transfer time.
The SP implodes under the magnetic pressure between 50 ns and 170 ns (see Fig. 3). This implosion time is about the Alfv\'en time $t_{A}\equiv r_{0}/v_{A}=\left(2r_{0}/I_{1}\right) \sqrt{\pi \lambda _{1}/\mu _{0}}$ (= 140 ns for $r_{0}=$ \SI{2.2}{mm}, $I_{1}=$ \SI{20}{kA}, and  $\lambda_{1}=$ \SI{1.6e-7}{kg/m}). 
The implosion leads to an increase of $R_1$, $dL/dt$, and  of $Z_1$ up to \SI{35}{m\ohm} at \SI{165}{ns} (Fig. \ref{fig:model} a).
Fast ionization of the LDP occurs around $t = $ \SI{130}{ns}. During the 20 ns ionization time, $N/n$  decreases by 2-3 orders of magnitude.  Subsequently,  $R_2$ decreases from \SI{2}{\ohm} to $\approx$\SI{10}{m\ohm}, below $Z_{1}$. 
This drastic decrease of the LDP resistance induces the current switching. The current transfer-time is $\tau =L/(Z_1+R_2)$.
As long as $R_2$ is large, $\tau$  is short and $I_2/I$ follows closely the impedance ratio $Z_{1}/(Z_{1}+R_{2})$.
However, as $R_{2}$ becomes small, $\tau$  increases so that the current partitioning $I_2/I$	lags behind the impedance ratio by $\approx$ \SI{25}{ns}, consistent with the experimental observations. 
Note that  $\tau$ is considerably shorter than the collisionless transfer time $L/(dL/dt)$, since the appropriate magnetic Reynolds number $(dL/dt)/R_1$ is smaller than unity.

We showed that in the presence of a \ac{LDP} there arises a competition  between the \ac{SP} and the \ac{LDP} for the discharge current.  
In the present experiment, the fast current redistribution can be explained by a drop of the LDP resistance $R_{2}$.
Similarly, the current may transfer if $Z_{1}$ increases to $Z_1 > R_{2}$, due to a fast compression of the SP, or an increase of $R_1$, caused by a large resistivity or a small areal cross section \cite{waisman2008resistance}.	
Consistent with this picture, MHD simulations performed in MACH2 \cite{peterkin1998mach2} show that in a setup with a very high resistance $R_2$, resulting from the lack of ionization in MACH2, the entire discharge current flows in the SP, throughout the stagnation. However, if the SP impedance $Z_1$ rises due to a forced compression of the SP to a  (possibly unrealistic) small radius, a fast current transfer to a  resistive LDP can be seen in the simulation.

Note that at locations farther away from the cathode, a gradually changing behavior is found: at z = 3 and \SI{5}{mm}, the discharge current is never observed to flow at small radii \cite{stollberg2019phd}, consistent with the absence of significant current in the stagnating plasma seen in  previous gas-puff experiments \cite{maron2013, rosenzweig2020pop, rosenzweig2017jinst, maron2020tutorial}.  
A larger abundance of peripheral gas further away from the cathode due to a divergent initial gas distribution (see Fig. \ref{fig:sketch}) results in an earlier diffusion of the current into the LDP forming a current that is \textit{trailing} the imploding plasma. This trailing current, that flows continuously at radii much larger than the SP radius in regions with significant abundance of 'trailing mass', has been observed previously \cite{rosenzweig2017jinst, rosenzweig2020pop, aybar2022, mikitchuk2019, maron2020tutorial} and was also indicated indirectly by imaging techniques and probe measurements \cite{lebedev2002snowplow, lebedev2005, waisman2004inductance, burdiak2013current, cuneo2005, hall2006}.

\textit{Conclusions - }
The fast current-escape from the \ac{SP} to much larger radii observed in the present work close to the cathode  is most likely related to the previously reported phenomenon of low current at the stagnation of \zpinches  \cite{maron2020tutorial, maron2013, rosenzweig2017jinst, rosenzweig2020pop, ivanov2015faraday},
 trailing current 
\cite{lebedev2002snowplow, lebedev2005, hall2006, burdiak2013current, waisman2004inductance, rosenzweig2017jinst, rosenzweig2020pop, aybar2022, mikitchuk2019, maron2020tutorial, cuneo2005}, 
and possible 'restriking' currents in wire-array \zpinches \cite{lebedev2002snowplow, lebedev2005, hall2006}. These phenomena might be different manifestations of similar mechanisms, governed by the interplay of advection and diffusion,  determined by the specific plasma and B-field parameters in these experiments. 
The fast current redistribution might also bear relevance to general pulsed-power systems, such as plasma switches \cite{engelbrecht2018pos, black2000pos, weber1995pos} or transmission lines \cite{gomez2017currentLoss, bennett2021transmissionLines}, where undesired plasmas  of different sources, such as electrodes
\cite{gomez2017currentLoss, bennett2021transmissionLines}, instabilities \cite{awe2013instability, giulliani2014temperatures}, evaporation and trailing mass in solid loads \cite{mitrofanov2017trailing, lebedev2002snowplow, lebedev2005, burdiak2013current, hall2006, waisman2004inductance, cuneo2005}, boundary layers in gas loads \cite{stollberg2019phd}, and unimploded plasma in \zpinches \cite{mikitchuk2019}, 
may affect the system operation.  
Further  investigations of the current switching should help  controlling it,   achieving higher efficiency in current-pulse transmission, and producing higher energy-density plasmas \cite{mikitchuk2019}.
Due to the fundamental aspects of this phenomenon it can also be relevant for  investigations of space plasmas \cite{camposRozo2019solarBfield, jarboe2019solarBfield}, and the evolution of the current distribution and energy balance in fusion plasmas \cite{maron2013, gomez2014maglif, sinars2020review}.

\begin{acknowledgments}	
	The authors acknowledge enlightening discussions with  A. Velikovich, V. Tangri, and Ya. E. Krasik,
	and are grateful to Brent Jones for valuable advise
	U. Shumlak gratefully acknowledges support of the Erna and Jakob Michael Visiting Professorship at the Weizmann Institute of Science.
	This work was supported in part 
	by the Cornell Multi-University Center for High Energy Density Science (USA), 
	by NSF-BSF (USA-Israel), 
	by the DOE/NNSA via the Naval Research Laboratory (USA), 
	by Sandia National Laboratories (USA), 
	by the Israel Science Foundation, 
	by the German Science Foundation (DFG TR18),
	by NNSA 83228-10966 [Prime No. DOE (NNSA) DE-NA0003764], and
	by NSF PHY-1805316.
\end{acknowledgments}


\begin{thebibliography}{55}%
	\makeatletter
	\providecommand \@ifxundefined [1]{%
		\@ifx{#1\undefined}
	}%
	\providecommand \@ifnum [1]{%
		\ifnum #1\expandafter \@firstoftwo
		\else \expandafter \@secondoftwo
		\fi
	}%
	\providecommand \@ifx [1]{%
		\ifx #1\expandafter \@firstoftwo
		\else \expandafter \@secondoftwo
		\fi
	}%
	\providecommand \natexlab [1]{#1}%
	\providecommand \enquote  [1]{``#1''}%
	\providecommand \bibnamefont  [1]{#1}%
	\providecommand \bibfnamefont [1]{#1}%
	\providecommand \citenamefont [1]{#1}%
	\providecommand \href@noop [0]{\@secondoftwo}%
	\providecommand \href [0]{\begingroup \@sanitize@url \@href}%
	\providecommand \@href[1]{\@@startlink{#1}\@@href}%
	\providecommand \@@href[1]{\endgroup#1\@@endlink}%
	\providecommand \@sanitize@url [0]{\catcode `\\12\catcode `\$12\catcode
		`\&12\catcode `\#12\catcode `\^12\catcode `\_12\catcode `\%12\relax}%
	\providecommand \@@startlink[1]{}%
	\providecommand \@@endlink[0]{}%
	\providecommand \url  [0]{\begingroup\@sanitize@url \@url }%
	\providecommand \@url [1]{\endgroup\@href {#1}{\urlprefix }}%
	\providecommand \urlprefix  [0]{URL }%
	\providecommand \Eprint [0]{\href }%
	\providecommand \doibase [0]{https://doi.org/}%
	\providecommand \selectlanguage [0]{\@gobble}%
	\providecommand \bibinfo  [0]{\@secondoftwo}%
	\providecommand \bibfield  [0]{\@secondoftwo}%
	\providecommand \translation [1]{[#1]}%
	\providecommand \BibitemOpen [0]{}%
	\providecommand \bibitemStop [0]{}%
	\providecommand \bibitemNoStop [0]{.\EOS\space}%
	\providecommand \EOS [0]{\spacefactor3000\relax}%
	\providecommand \BibitemShut  [1]{\csname bibitem#1\endcsname}%
	\let\auto@bib@innerbib\@empty
	\bibitem [{\citenamefont {Rakov}\ and\ \citenamefont
		{Uman}(2004)}]{burt2004lightning}%
	\BibitemOpen
	\bibfield  {author} {\bibinfo {author} {\bibfnamefont {V.~A.}\ \bibnamefont
			{Rakov}}\ and\ \bibinfo {author} {\bibfnamefont {M.~A.}\ \bibnamefont
			{Uman}},\ }\bibfield  {title} {\bibinfo {title} {Lightning physics and
			effects},\ }\href {https://doi.org/10.1256/wea.168/03} {\bibfield  {journal}
		{\bibinfo  {journal} {Cambridge University Press}\ }\textbf {\bibinfo
			{volume} {59}},\ \bibinfo {pages} {109} (\bibinfo {year} {2004})}\BibitemShut
	{NoStop}%
	\bibitem [{\citenamefont {Chen}(2017)}]{chen2017sun}%
	\BibitemOpen
	\bibfield  {author} {\bibinfo {author} {\bibfnamefont {J.}~\bibnamefont
			{Chen}},\ }\bibfield  {title} {\bibinfo {title} {Physics of erupting solar
			flux ropes: Coronal mass ejections: Recent advances in theory and
			observation},\ }\href {https://doi.org/10.1063/1.4993929} {\bibfield
		{journal} {\bibinfo  {journal} {Physics of Plasmas}\ }\textbf {\bibinfo
			{volume} {24}},\ \bibinfo {pages} {090501} (\bibinfo {year} {2017})},\
	\Eprint {https://arxiv.org/abs/https://doi.org/10.1063/1.4993929}
	{https://doi.org/10.1063/1.4993929} \BibitemShut {NoStop}%
	\bibitem [{\citenamefont {Tzeferacos}\ \emph {et~al.}(2018)\citenamefont
		{Tzeferacos}, \citenamefont {Rigby}, \citenamefont {Bott}, \citenamefont
		{Bell}, \citenamefont {Bingham}, \citenamefont {Casner}, \citenamefont
		{Cattaneo}, \citenamefont {Churazov}, \citenamefont {Emig}, \citenamefont
		{Fiuza}, \citenamefont {Forest}, \citenamefont {Foster}, \citenamefont
		{Graziani}, \citenamefont {Katz}, \citenamefont {Koenig}, \citenamefont {Li},
		\citenamefont {Meinecke}, \citenamefont {Petrasso}, \citenamefont {Park},
		\citenamefont {Remington}, \citenamefont {Ross}, \citenamefont {Ryu},
		\citenamefont {Ryutov}, \citenamefont {White}, \citenamefont {Reville},
		\citenamefont {Miniati}, \citenamefont {Schekochihin}, \citenamefont {Lamb},
		\citenamefont {Froula},\ and\ \citenamefont
		{Gregori}}]{tzeferacos2018dynamo}%
	\BibitemOpen
	\bibfield  {author} {\bibinfo {author} {\bibfnamefont {P.}~\bibnamefont
			{Tzeferacos}}, \bibinfo {author} {\bibfnamefont {A.}~\bibnamefont {Rigby}},
		\bibinfo {author} {\bibfnamefont {A.~F.~A.}\ \bibnamefont {Bott}}, \bibinfo
		{author} {\bibfnamefont {A.~R.}\ \bibnamefont {Bell}}, \bibinfo {author}
		{\bibfnamefont {R.}~\bibnamefont {Bingham}}, \bibinfo {author} {\bibfnamefont
			{A.}~\bibnamefont {Casner}}, \bibinfo {author} {\bibfnamefont
			{F.}~\bibnamefont {Cattaneo}}, \bibinfo {author} {\bibfnamefont {E.~M.}\
			\bibnamefont {Churazov}}, \bibinfo {author} {\bibfnamefont {J.}~\bibnamefont
			{Emig}}, \bibinfo {author} {\bibfnamefont {F.}~\bibnamefont {Fiuza}},
		\bibinfo {author} {\bibfnamefont {C.~B.}\ \bibnamefont {Forest}}, \bibinfo
		{author} {\bibfnamefont {J.}~\bibnamefont {Foster}}, \bibinfo {author}
		{\bibfnamefont {C.}~\bibnamefont {Graziani}}, \bibinfo {author}
		{\bibfnamefont {J.}~\bibnamefont {Katz}}, \bibinfo {author} {\bibfnamefont
			{M.}~\bibnamefont {Koenig}}, \bibinfo {author} {\bibfnamefont {C.-K.}\
			\bibnamefont {Li}}, \bibinfo {author} {\bibfnamefont {J.}~\bibnamefont
			{Meinecke}}, \bibinfo {author} {\bibfnamefont {R.}~\bibnamefont {Petrasso}},
		\bibinfo {author} {\bibfnamefont {H.-S.}\ \bibnamefont {Park}}, \bibinfo
		{author} {\bibfnamefont {B.~A.}\ \bibnamefont {Remington}}, \bibinfo {author}
		{\bibfnamefont {J.~S.}\ \bibnamefont {Ross}}, \bibinfo {author}
		{\bibfnamefont {D.}~\bibnamefont {Ryu}}, \bibinfo {author} {\bibfnamefont
			{D.}~\bibnamefont {Ryutov}}, \bibinfo {author} {\bibfnamefont {T.~G.}\
			\bibnamefont {White}}, \bibinfo {author} {\bibfnamefont {B.}~\bibnamefont
			{Reville}}, \bibinfo {author} {\bibfnamefont {F.}~\bibnamefont {Miniati}},
		\bibinfo {author} {\bibfnamefont {A.~A.}\ \bibnamefont {Schekochihin}},
		\bibinfo {author} {\bibfnamefont {D.~Q.}\ \bibnamefont {Lamb}}, \bibinfo
		{author} {\bibfnamefont {D.~H.}\ \bibnamefont {Froula}},\ and\ \bibinfo
		{author} {\bibfnamefont {G.}~\bibnamefont {Gregori}},\ }\bibfield  {title}
	{\bibinfo {title} {Laboratory evidence of dynamo amplification of magnetic
			fields in a turbulent plasma},\ }\href
	{https://doi.org/10.1038/s41467-018-02953-2} {\bibfield  {journal} {\bibinfo
			{journal} {Nature Communications}\ }\textbf {\bibinfo {volume} {9}},\
		\bibinfo {pages} {591} (\bibinfo {year} {2018})}\BibitemShut {NoStop}%
	\bibitem [{\citenamefont {Kikuchi}(2010)}]{kikuchi2010fusionReview}%
	\BibitemOpen
	\bibfield  {author} {\bibinfo {author} {\bibfnamefont {M.}~\bibnamefont
			{Kikuchi}},\ }\bibfield  {title} {\bibinfo {title} {A review of fusion and
			tokamak research towards steady-state operation: A jaea contribution},\
	}\href {https://doi.org/https://www.mdpi.com/1996-1073/3/11/1741} {\bibfield
		{journal} {\bibinfo  {journal} {Energies}\ }\textbf {\bibinfo {volume} {3}},\
		\bibinfo {pages} {1741} (\bibinfo {year} {2010})}\BibitemShut {NoStop}%
	\bibitem [{\citenamefont {Sinars}\ \emph {et~al.}(2020)\citenamefont {Sinars},
		\citenamefont {Sweeney} \emph {et~al.}}]{sinars2020review}%
	\BibitemOpen
	\bibfield  {author} {\bibinfo {author} {\bibfnamefont {D.~B.}\ \bibnamefont
			{Sinars}}, \bibinfo {author} {\bibfnamefont {M.~A.}\ \bibnamefont {Sweeney}},
		\emph {et~al.},\ }\bibfield  {title} {\bibinfo {title} {Review of pulsed
			power-driven high energy density physics research on z at sandia},\ }\href
	{https://doi.org/10.1063/5.0007476} {\bibfield  {journal} {\bibinfo
			{journal} {Physics of Plasmas}\ }\textbf {\bibinfo {volume} {27}},\ \bibinfo
		{pages} {070501} (\bibinfo {year} {2020})},\ \Eprint
	{https://arxiv.org/abs/https://doi.org/10.1063/5.0007476}
	{https://doi.org/10.1063/5.0007476} \BibitemShut {NoStop}%
	\bibitem [{\citenamefont {Ryutov}\ \emph {et~al.}(2000)\citenamefont {Ryutov},
		\citenamefont {Derzon},\ and\ \citenamefont {Matzen}}]{ryutov2000review}%
	\BibitemOpen
	\bibfield  {author} {\bibinfo {author} {\bibfnamefont {D.~D.}\ \bibnamefont
			{Ryutov}}, \bibinfo {author} {\bibfnamefont {M.~S.}\ \bibnamefont {Derzon}},\
		and\ \bibinfo {author} {\bibfnamefont {M.~K.}\ \bibnamefont {Matzen}},\
	}\bibfield  {title} {\bibinfo {title} {The physics of fast $z$ pinches},\
	}\href {https://doi.org/10.1103/RevModPhys.72.167} {\bibfield  {journal}
		{\bibinfo  {journal} {Rev. Mod. Phys.}\ }\textbf {\bibinfo {volume} {72}},\
		\bibinfo {pages} {167} (\bibinfo {year} {2000})}\BibitemShut {NoStop}%
	\bibitem [{\citenamefont {Giuliani}\ and\ \citenamefont
		{Commisso}(2015)}]{giuliani2015review}%
	\BibitemOpen
	\bibfield  {author} {\bibinfo {author} {\bibfnamefont {J.~L.}\ \bibnamefont
			{Giuliani}}\ and\ \bibinfo {author} {\bibfnamefont {R.~J.}\ \bibnamefont
			{Commisso}},\ }\bibfield  {title} {\bibinfo {title} {A review of the
			gas-puff$z$-pinch as an x-ray and neutron source},\ }\href
	{https://doi.org/10.1109/TPS.2015.2451157} {\bibfield  {journal} {\bibinfo
			{journal} {IEEE Transactions on Plasma Science}\ }\textbf {\bibinfo {volume}
			{43}},\ \bibinfo {pages} {2385} (\bibinfo {year} {2015})}\BibitemShut
	{NoStop}%
	\bibitem [{\citenamefont {Slutz}\ and\ \citenamefont
		{Vesey}(2012)}]{slutz2012prl}%
	\BibitemOpen
	\bibfield  {author} {\bibinfo {author} {\bibfnamefont {S.~A.}\ \bibnamefont
			{Slutz}}\ and\ \bibinfo {author} {\bibfnamefont {R.~A.}\ \bibnamefont
			{Vesey}},\ }\bibfield  {title} {\bibinfo {title} {High-gain magnetized
			inertial fusion},\ }\href {https://doi.org/10.1103/PhysRevLett.108.025003}
	{\bibfield  {journal} {\bibinfo  {journal} {Phys. Rev. Lett.}\ }\textbf
		{\bibinfo {volume} {108}},\ \bibinfo {pages} {025003} (\bibinfo {year}
		{2012})}\BibitemShut {NoStop}%
	\bibitem [{\citenamefont {Gomez}\ \emph
		{et~al.}(2014{\natexlab{a}})\citenamefont {Gomez}, \citenamefont {Slutz},
		\citenamefont {Sefkow}, \citenamefont {Sinars}, \citenamefont {Hahn},
		\citenamefont {Hansen}, \citenamefont {Harding}, \citenamefont {Knapp},
		\citenamefont {Schmit}, \citenamefont {Jennings}, \citenamefont {Awe},
		\citenamefont {Geissel}, \citenamefont {Rovang}, \citenamefont {Chandler},
		\citenamefont {Cooper}, \citenamefont {Cuneo}, \citenamefont
		{Harvey-Thompson}, \citenamefont {Herrmann}, \citenamefont {Hess},
		\citenamefont {Johns}, \citenamefont {Lamppa}, \citenamefont {Martin},
		\citenamefont {McBride}, \citenamefont {Peterson}, \citenamefont {Porter},
		\citenamefont {Robertson}, \citenamefont {Rochau}, \citenamefont {Ruiz},
		\citenamefont {Savage}, \citenamefont {Smith}, \citenamefont {Stygar},\ and\
		\citenamefont {Vesey}}]{gomez2014maglif}%
	\BibitemOpen
	\bibfield  {author} {\bibinfo {author} {\bibfnamefont {M.~R.}\ \bibnamefont
			{Gomez}}, \bibinfo {author} {\bibfnamefont {S.~A.}\ \bibnamefont {Slutz}},
		\bibinfo {author} {\bibfnamefont {A.~B.}\ \bibnamefont {Sefkow}}, \bibinfo
		{author} {\bibfnamefont {D.~B.}\ \bibnamefont {Sinars}}, \bibinfo {author}
		{\bibfnamefont {K.~D.}\ \bibnamefont {Hahn}}, \bibinfo {author}
		{\bibfnamefont {S.~B.}\ \bibnamefont {Hansen}}, \bibinfo {author}
		{\bibfnamefont {E.~C.}\ \bibnamefont {Harding}}, \bibinfo {author}
		{\bibfnamefont {P.~F.}\ \bibnamefont {Knapp}}, \bibinfo {author}
		{\bibfnamefont {P.~F.}\ \bibnamefont {Schmit}}, \bibinfo {author}
		{\bibfnamefont {C.~A.}\ \bibnamefont {Jennings}}, \bibinfo {author}
		{\bibfnamefont {T.~J.}\ \bibnamefont {Awe}}, \bibinfo {author} {\bibfnamefont
			{M.}~\bibnamefont {Geissel}}, \bibinfo {author} {\bibfnamefont {D.~C.}\
			\bibnamefont {Rovang}}, \bibinfo {author} {\bibfnamefont {G.~A.}\
			\bibnamefont {Chandler}}, \bibinfo {author} {\bibfnamefont {G.~W.}\
			\bibnamefont {Cooper}}, \bibinfo {author} {\bibfnamefont {M.~E.}\
			\bibnamefont {Cuneo}}, \bibinfo {author} {\bibfnamefont {A.~J.}\ \bibnamefont
			{Harvey-Thompson}}, \bibinfo {author} {\bibfnamefont {M.~C.}\ \bibnamefont
			{Herrmann}}, \bibinfo {author} {\bibfnamefont {M.~H.}\ \bibnamefont {Hess}},
		\bibinfo {author} {\bibfnamefont {O.}~\bibnamefont {Johns}}, \bibinfo
		{author} {\bibfnamefont {D.~C.}\ \bibnamefont {Lamppa}}, \bibinfo {author}
		{\bibfnamefont {M.~R.}\ \bibnamefont {Martin}}, \bibinfo {author}
		{\bibfnamefont {R.~D.}\ \bibnamefont {McBride}}, \bibinfo {author}
		{\bibfnamefont {K.~J.}\ \bibnamefont {Peterson}}, \bibinfo {author}
		{\bibfnamefont {J.~L.}\ \bibnamefont {Porter}}, \bibinfo {author}
		{\bibfnamefont {G.~K.}\ \bibnamefont {Robertson}}, \bibinfo {author}
		{\bibfnamefont {G.~A.}\ \bibnamefont {Rochau}}, \bibinfo {author}
		{\bibfnamefont {C.~L.}\ \bibnamefont {Ruiz}}, \bibinfo {author}
		{\bibfnamefont {M.~E.}\ \bibnamefont {Savage}}, \bibinfo {author}
		{\bibfnamefont {I.~C.}\ \bibnamefont {Smith}}, \bibinfo {author}
		{\bibfnamefont {W.~A.}\ \bibnamefont {Stygar}},\ and\ \bibinfo {author}
		{\bibfnamefont {R.~A.}\ \bibnamefont {Vesey}},\ }\bibfield  {title} {\bibinfo
		{title} {Experimental demonstration of fusion-relevant conditions in
			magnetized liner inertial fusion},\ }\href
	{https://doi.org/10.1103/PhysRevLett.113.155003} {\bibfield  {journal}
		{\bibinfo  {journal} {Phys. Rev. Lett.}\ }\textbf {\bibinfo {volume} {113}},\
		\bibinfo {pages} {155003} (\bibinfo {year} {2014}{\natexlab{a}})}\BibitemShut
	{NoStop}%
	\bibitem [{\citenamefont {Maron}(2020)}]{maron2020tutorial}%
	\BibitemOpen
	\bibfield  {author} {\bibinfo {author} {\bibfnamefont {Y.}~\bibnamefont
			{Maron}},\ }\bibfield  {title} {\bibinfo {title} {Experimental determination
			of the thermal, turbulent, and rotational ion motion and magnetic field
			profiles in imploding plasmas},\ }\href {https://doi.org/10.1063/5.0009432}
	{\bibfield  {journal} {\bibinfo  {journal} {Physics of Plasmas}\ }\textbf
		{\bibinfo {volume} {27}},\ \bibinfo {pages} {060901} (\bibinfo {year}
		{2020})},\ \Eprint {https://arxiv.org/abs/https://doi.org/10.1063/5.0009432}
	{https://doi.org/10.1063/5.0009432} \BibitemShut {NoStop}%
	\bibitem [{\citenamefont {Black}\ \emph {et~al.}(2000)\citenamefont {Black},
		\citenamefont {Commisso}, \citenamefont {Ottinger}, \citenamefont
		{Swanekamp},\ and\ \citenamefont {Weber}}]{black2000pos}%
	\BibitemOpen
	\bibfield  {author} {\bibinfo {author} {\bibfnamefont {D.~C.}\ \bibnamefont
			{Black}}, \bibinfo {author} {\bibfnamefont {R.~J.}\ \bibnamefont {Commisso}},
		\bibinfo {author} {\bibfnamefont {P.~F.}\ \bibnamefont {Ottinger}}, \bibinfo
		{author} {\bibfnamefont {S.~B.}\ \bibnamefont {Swanekamp}},\ and\ \bibinfo
		{author} {\bibfnamefont {B.~V.}\ \bibnamefont {Weber}},\ }\bibfield  {title}
	{\bibinfo {title} {Experimental determination of gap scaling in a plasma
			opening switch},\ }\href {https://doi.org/10.1063/1.1287914} {\bibfield
		{journal} {\bibinfo  {journal} {Physics of Plasmas}\ }\textbf {\bibinfo
			{volume} {7}},\ \bibinfo {pages} {3790} (\bibinfo {year} {2000})},\ \Eprint
	{https://arxiv.org/abs/https://doi.org/10.1063/1.1287914}
	{https://doi.org/10.1063/1.1287914} \BibitemShut {NoStop}%
	\bibitem [{\citenamefont {Engelbrecht}\ \emph {et~al.}(2018)\citenamefont
		{Engelbrecht}, \citenamefont {Ouart}, \citenamefont {Qi}, \citenamefont
		{de~Grouchy}, \citenamefont {Shelkovenko}, \citenamefont {Pikuz},
		\citenamefont {Banasek}, \citenamefont {Potter}, \citenamefont {Rocco},
		\citenamefont {Hammer}, \citenamefont {Kusse},\ and\ \citenamefont
		{Giuliani}}]{engelbrecht2018pos}%
	\BibitemOpen
	\bibfield  {author} {\bibinfo {author} {\bibfnamefont {J.~T.}\ \bibnamefont
			{Engelbrecht}}, \bibinfo {author} {\bibfnamefont {N.~D.}\ \bibnamefont
			{Ouart}}, \bibinfo {author} {\bibfnamefont {N.}~\bibnamefont {Qi}}, \bibinfo
		{author} {\bibfnamefont {P.~W.}\ \bibnamefont {de~Grouchy}}, \bibinfo
		{author} {\bibfnamefont {T.~A.}\ \bibnamefont {Shelkovenko}}, \bibinfo
		{author} {\bibfnamefont {S.~A.}\ \bibnamefont {Pikuz}}, \bibinfo {author}
		{\bibfnamefont {J.~T.}\ \bibnamefont {Banasek}}, \bibinfo {author}
		{\bibfnamefont {W.~M.}\ \bibnamefont {Potter}}, \bibinfo {author}
		{\bibfnamefont {S.~V.}\ \bibnamefont {Rocco}}, \bibinfo {author}
		{\bibfnamefont {D.~A.}\ \bibnamefont {Hammer}}, \bibinfo {author}
		{\bibfnamefont {B.~R.}\ \bibnamefont {Kusse}},\ and\ \bibinfo {author}
		{\bibfnamefont {J.~L.}\ \bibnamefont {Giuliani}},\ }\bibfield  {title}
	{\bibinfo {title} {Enhancing the x-ray output of a single-wire explosion with
			a gas-puff based plasma opening switch},\ }\href
	{https://doi.org/10.1063/1.5019378} {\bibfield  {journal} {\bibinfo
			{journal} {Physics of Plasmas}\ }\textbf {\bibinfo {volume} {25}},\ \bibinfo
		{pages} {022704} (\bibinfo {year} {2018})},\ \Eprint
	{https://arxiv.org/abs/https://doi.org/10.1063/1.5019378}
	{https://doi.org/10.1063/1.5019378} \BibitemShut {NoStop}%
	\bibitem [{\citenamefont {Weber}\ \emph {et~al.}(1995)\citenamefont {Weber},
		\citenamefont {Commisso}, \citenamefont {Goodrich}, \citenamefont
		{Grossmann}, \citenamefont {Hinshelwood}, \citenamefont {Ottinger},\ and\
		\citenamefont {Swanekamp}}]{weber1995pos}%
	\BibitemOpen
	\bibfield  {author} {\bibinfo {author} {\bibfnamefont {B.~V.}\ \bibnamefont
			{Weber}}, \bibinfo {author} {\bibfnamefont {R.~J.}\ \bibnamefont {Commisso}},
		\bibinfo {author} {\bibfnamefont {P.~J.}\ \bibnamefont {Goodrich}}, \bibinfo
		{author} {\bibfnamefont {J.~M.}\ \bibnamefont {Grossmann}}, \bibinfo {author}
		{\bibfnamefont {D.~D.}\ \bibnamefont {Hinshelwood}}, \bibinfo {author}
		{\bibfnamefont {P.~F.}\ \bibnamefont {Ottinger}},\ and\ \bibinfo {author}
		{\bibfnamefont {S.~B.}\ \bibnamefont {Swanekamp}},\ }\bibfield  {title}
	{\bibinfo {title} {Plasma opening switch conduction scaling},\ }\href
	{https://doi.org/10.1063/1.871018} {\bibfield  {journal} {\bibinfo  {journal}
			{Physics of Plasmas}\ }\textbf {\bibinfo {volume} {2}},\ \bibinfo {pages}
		{3893} (\bibinfo {year} {1995})},\ \Eprint
	{https://arxiv.org/abs/https://doi.org/10.1063/1.871018}
	{https://doi.org/10.1063/1.871018} \BibitemShut {NoStop}%
	\bibitem [{\citenamefont {Gomez}\ \emph {et~al.}(2017)\citenamefont {Gomez},
		\citenamefont {Gilgenbach}, \citenamefont {Cuneo}, \citenamefont {Jennings},
		\citenamefont {McBride}, \citenamefont {Waisman}, \citenamefont {Hutsel},
		\citenamefont {Stygar}, \citenamefont {Rose},\ and\ \citenamefont
		{Maron}}]{gomez2017currentLoss}%
	\BibitemOpen
	\bibfield  {author} {\bibinfo {author} {\bibfnamefont {M.~R.}\ \bibnamefont
			{Gomez}}, \bibinfo {author} {\bibfnamefont {R.~M.}\ \bibnamefont
			{Gilgenbach}}, \bibinfo {author} {\bibfnamefont {M.~E.}\ \bibnamefont
			{Cuneo}}, \bibinfo {author} {\bibfnamefont {C.~A.}\ \bibnamefont {Jennings}},
		\bibinfo {author} {\bibfnamefont {R.~D.}\ \bibnamefont {McBride}}, \bibinfo
		{author} {\bibfnamefont {E.~M.}\ \bibnamefont {Waisman}}, \bibinfo {author}
		{\bibfnamefont {B.~T.}\ \bibnamefont {Hutsel}}, \bibinfo {author}
		{\bibfnamefont {W.~A.}\ \bibnamefont {Stygar}}, \bibinfo {author}
		{\bibfnamefont {D.~V.}\ \bibnamefont {Rose}},\ and\ \bibinfo {author}
		{\bibfnamefont {Y.}~\bibnamefont {Maron}},\ }\bibfield  {title} {\bibinfo
		{title} {Experimental study of current loss and plasma formation in the $z$
			machine post-hole convolute},\ }\href
	{https://doi.org/10.1103/PhysRevAccelBeams.20.010401} {\bibfield  {journal}
		{\bibinfo  {journal} {Phys. Rev. Accel. Beams}\ }\textbf {\bibinfo {volume}
			{20}},\ \bibinfo {pages} {010401} (\bibinfo {year} {2017})}\BibitemShut
	{NoStop}%
	\bibitem [{\citenamefont {Bennett}\ \emph {et~al.}(2021)\citenamefont
		{Bennett}, \citenamefont {Welch}, \citenamefont {Laity}, \citenamefont
		{Rose},\ and\ \citenamefont {Cuneo}}]{bennett2021transmissionLines}%
	\BibitemOpen
	\bibfield  {author} {\bibinfo {author} {\bibfnamefont {N.}~\bibnamefont
			{Bennett}}, \bibinfo {author} {\bibfnamefont {D.~R.}\ \bibnamefont {Welch}},
		\bibinfo {author} {\bibfnamefont {G.}~\bibnamefont {Laity}}, \bibinfo
		{author} {\bibfnamefont {D.~V.}\ \bibnamefont {Rose}},\ and\ \bibinfo
		{author} {\bibfnamefont {M.~E.}\ \bibnamefont {Cuneo}},\ }\bibfield  {title}
	{\bibinfo {title} {Magnetized particle transport in multi-ma accelerators},\
	}\href {https://doi.org/10.1103/PhysRevAccelBeams.24.060401} {\bibfield
		{journal} {\bibinfo  {journal} {Phys. Rev. Accel. Beams}\ }\textbf {\bibinfo
			{volume} {24}},\ \bibinfo {pages} {060401} (\bibinfo {year}
		{2021})}\BibitemShut {NoStop}%
	\bibitem [{\citenamefont {Lebedev}\ \emph {et~al.}(2019)\citenamefont
		{Lebedev}, \citenamefont {Frank},\ and\ \citenamefont
		{Ryutov}}]{lebedev2019astrophysics}%
	\BibitemOpen
	\bibfield  {author} {\bibinfo {author} {\bibfnamefont {S.~V.}\ \bibnamefont
			{Lebedev}}, \bibinfo {author} {\bibfnamefont {A.}~\bibnamefont {Frank}},\
		and\ \bibinfo {author} {\bibfnamefont {D.~D.}\ \bibnamefont {Ryutov}},\
	}\bibfield  {title} {\bibinfo {title} {Exploring astrophysics-relevant
			magnetohydrodynamics with pulsed-power laboratory facilities},\ }\href
	{https://doi.org/10.1103/RevModPhys.91.025002} {\bibfield  {journal}
		{\bibinfo  {journal} {Rev. Mod. Phys.}\ }\textbf {\bibinfo {volume} {91}},\
		\bibinfo {pages} {025002} (\bibinfo {year} {2019})}\BibitemShut {NoStop}%
	\bibitem [{\citenamefont {Alumot}\ \emph {et~al.}(2019)\citenamefont {Alumot},
		\citenamefont {Kroupp}, \citenamefont {Stambulchik}, \citenamefont
		{Starobinets}, \citenamefont {Uschmann},\ and\ \citenamefont
		{Maron}}]{alumot2019ionTemperature}%
	\BibitemOpen
	\bibfield  {author} {\bibinfo {author} {\bibfnamefont {D.}~\bibnamefont
			{Alumot}}, \bibinfo {author} {\bibfnamefont {E.}~\bibnamefont {Kroupp}},
		\bibinfo {author} {\bibfnamefont {E.}~\bibnamefont {Stambulchik}}, \bibinfo
		{author} {\bibfnamefont {A.}~\bibnamefont {Starobinets}}, \bibinfo {author}
		{\bibfnamefont {I.}~\bibnamefont {Uschmann}},\ and\ \bibinfo {author}
		{\bibfnamefont {Y.}~\bibnamefont {Maron}},\ }\bibfield  {title} {\bibinfo
		{title} {Determination of the ion temperature in a high-energy-density plasma
			using the {Stark} effect},\ }\href
	{https://doi.org/10.1103/PhysRevLett.122.095001} {\bibfield  {journal}
		{\bibinfo  {journal} {Phys. Rev. Lett.}\ }\textbf {\bibinfo {volume} {122}},\
		\bibinfo {pages} {095001} (\bibinfo {year} {2019})}\BibitemShut {NoStop}%
	\bibitem [{\citenamefont {Kroupp}\ \emph {et~al.}(2018)\citenamefont {Kroupp},
		\citenamefont {Stambulchik}, \citenamefont {Starobinets}, \citenamefont
		{Osin}, \citenamefont {Fisher}, \citenamefont {Alumot}, \citenamefont
		{Maron}, \citenamefont {Davidovits}, \citenamefont {Fisch},\ and\
		\citenamefont {Fruchtman}}]{kroupp2018turbulent}%
	\BibitemOpen
	\bibfield  {author} {\bibinfo {author} {\bibfnamefont {E.}~\bibnamefont
			{Kroupp}}, \bibinfo {author} {\bibfnamefont {E.}~\bibnamefont {Stambulchik}},
		\bibinfo {author} {\bibfnamefont {A.}~\bibnamefont {Starobinets}}, \bibinfo
		{author} {\bibfnamefont {D.}~\bibnamefont {Osin}}, \bibinfo {author}
		{\bibfnamefont {V.~I.}\ \bibnamefont {Fisher}}, \bibinfo {author}
		{\bibfnamefont {D.}~\bibnamefont {Alumot}}, \bibinfo {author} {\bibfnamefont
			{Y.}~\bibnamefont {Maron}}, \bibinfo {author} {\bibfnamefont
			{S.}~\bibnamefont {Davidovits}}, \bibinfo {author} {\bibfnamefont {N.~J.}\
			\bibnamefont {Fisch}},\ and\ \bibinfo {author} {\bibfnamefont
			{A.}~\bibnamefont {Fruchtman}},\ }\bibfield  {title} {\bibinfo {title}
		{Turbulent stagnation in a $z$-pinch plasma},\ }\href
	{https://doi.org/10.1103/PhysRevE.97.013202} {\bibfield  {journal} {\bibinfo
			{journal} {Phys. Rev. E}\ }\textbf {\bibinfo {volume} {97}},\ \bibinfo
		{pages} {013202} (\bibinfo {year} {2018})}\BibitemShut {NoStop}%
	\bibitem [{\citenamefont {Lebedev}\ \emph {et~al.}(2004)\citenamefont
		{Lebedev}, \citenamefont {Ampleford}, \citenamefont {Bland}, \citenamefont
		{Bott}, \citenamefont {Chittenden}, \citenamefont {Jennings}, \citenamefont
		{Haines}, \citenamefont {Palmer},\ and\ \citenamefont
		{Rapley}}]{lebedev2004imperialCollege}%
	\BibitemOpen
	\bibfield  {author} {\bibinfo {author} {\bibfnamefont {S.}~\bibnamefont
			{Lebedev}}, \bibinfo {author} {\bibfnamefont {D.}~\bibnamefont {Ampleford}},
		\bibinfo {author} {\bibfnamefont {S.}~\bibnamefont {Bland}}, \bibinfo
		{author} {\bibfnamefont {S.}~\bibnamefont {Bott}}, \bibinfo {author}
		{\bibfnamefont {J.}~\bibnamefont {Chittenden}}, \bibinfo {author}
		{\bibfnamefont {C.}~\bibnamefont {Jennings}}, \bibinfo {author}
		{\bibfnamefont {M.}~\bibnamefont {Haines}}, \bibinfo {author} {\bibfnamefont
			{J.}~\bibnamefont {Palmer}},\ and\ \bibinfo {author} {\bibfnamefont
			{J.}~\bibnamefont {Rapley}},\ }\bibfield  {title} {\bibinfo {title}
		{Implosion dynamics of wire array z-pinches: experiments at imperial
			college},\ }\href {https://doi.org/10.1088/0029-5515/44/12/s12} {\bibfield
		{journal} {\bibinfo  {journal} {Nucl. Fusion}\ }\textbf {\bibinfo {volume}
			{44}},\ \bibinfo {pages} {S215} (\bibinfo {year} {2004})}\BibitemShut
	{NoStop}%
	\bibitem [{\citenamefont {Maron}\ \emph {et~al.}(2013)\citenamefont {Maron},
		\citenamefont {Starobinets}, \citenamefont {Fisher}, \citenamefont {Kroupp},
		\citenamefont {Osin}, \citenamefont {Fisher}, \citenamefont {Deeney},
		\citenamefont {Coverdale}, \citenamefont {Lepell}, \citenamefont {Yu},
		\citenamefont {Jennings}, \citenamefont {Cuneo}, \citenamefont {Herrmann},
		\citenamefont {Porter}, \citenamefont {Mehlhorn},\ and\ \citenamefont
		{Apruzese}}]{maron2013}%
	\BibitemOpen
	\bibfield  {author} {\bibinfo {author} {\bibfnamefont {Y.}~\bibnamefont
			{Maron}}, \bibinfo {author} {\bibfnamefont {A.}~\bibnamefont {Starobinets}},
		\bibinfo {author} {\bibfnamefont {V.~I.}\ \bibnamefont {Fisher}}, \bibinfo
		{author} {\bibfnamefont {E.}~\bibnamefont {Kroupp}}, \bibinfo {author}
		{\bibfnamefont {D.}~\bibnamefont {Osin}}, \bibinfo {author} {\bibfnamefont
			{A.}~\bibnamefont {Fisher}}, \bibinfo {author} {\bibfnamefont
			{C.}~\bibnamefont {Deeney}}, \bibinfo {author} {\bibfnamefont {C.~A.}\
			\bibnamefont {Coverdale}}, \bibinfo {author} {\bibfnamefont {P.~D.}\
			\bibnamefont {Lepell}}, \bibinfo {author} {\bibfnamefont {E.~P.}\
			\bibnamefont {Yu}}, \bibinfo {author} {\bibfnamefont {C.}~\bibnamefont
			{Jennings}}, \bibinfo {author} {\bibfnamefont {M.~E.}\ \bibnamefont {Cuneo}},
		\bibinfo {author} {\bibfnamefont {M.~C.}\ \bibnamefont {Herrmann}}, \bibinfo
		{author} {\bibfnamefont {J.~L.}\ \bibnamefont {Porter}}, \bibinfo {author}
		{\bibfnamefont {T.~A.}\ \bibnamefont {Mehlhorn}},\ and\ \bibinfo {author}
		{\bibfnamefont {J.~P.}\ \bibnamefont {Apruzese}},\ }\bibfield  {title}
	{\bibinfo {title} {Pressure and energy balance of stagnating plasmas in
			$z$-pinch experiments: Implications to current flow at stagnation},\ }\href
	{https://doi.org/10.1103/PhysRevLett.111.035001} {\bibfield  {journal}
		{\bibinfo  {journal} {Phys. Rev. Lett.}\ }\textbf {\bibinfo {volume} {111}},\
		\bibinfo {pages} {035001} (\bibinfo {year} {2013})}\BibitemShut {NoStop}%
	\bibitem [{\citenamefont {Doron}\ \emph {et~al.}(2014)\citenamefont {Doron},
		\citenamefont {Mikitchuk}, \citenamefont {Stollberg}, \citenamefont
		{Rosenzweig}, \citenamefont {Stambulchik}, \citenamefont {Kroupp},
		\citenamefont {Maron},\ and\ \citenamefont {Hammer}}]{doron2014bfield}%
	\BibitemOpen
	\bibfield  {author} {\bibinfo {author} {\bibfnamefont {R.}~\bibnamefont
			{Doron}}, \bibinfo {author} {\bibfnamefont {D.}~\bibnamefont {Mikitchuk}},
		\bibinfo {author} {\bibfnamefont {C.}~\bibnamefont {Stollberg}}, \bibinfo
		{author} {\bibfnamefont {G.}~\bibnamefont {Rosenzweig}}, \bibinfo {author}
		{\bibfnamefont {E.}~\bibnamefont {Stambulchik}}, \bibinfo {author}
		{\bibfnamefont {E.}~\bibnamefont {Kroupp}}, \bibinfo {author} {\bibfnamefont
			{Y.}~\bibnamefont {Maron}},\ and\ \bibinfo {author} {\bibfnamefont
			{D.}~\bibnamefont {Hammer}},\ }\bibfield  {title} {\bibinfo {title}
		{Determination of magnetic fields based on the zeeman effect in regimes
			inaccessible by zeeman-splitting spectroscopy},\ }\href
	{https://doi.org/https://doi.org/10.1016/j.hedp.2013.10.004} {\bibfield
		{journal} {\bibinfo  {journal} {High Energy Density Physics}\ }\textbf
		{\bibinfo {volume} {10}},\ \bibinfo {pages} {56 } (\bibinfo {year}
		{2014})}\BibitemShut {NoStop}%
	\bibitem [{\citenamefont {Davara}\ \emph {et~al.}(1998)\citenamefont {Davara},
		\citenamefont {Gregorian}, \citenamefont {Kroupp},\ and\ \citenamefont
		{Maron}}]{davara1998}%
	\BibitemOpen
	\bibfield  {author} {\bibinfo {author} {\bibfnamefont {G.}~\bibnamefont
			{Davara}}, \bibinfo {author} {\bibfnamefont {L.}~\bibnamefont {Gregorian}},
		\bibinfo {author} {\bibfnamefont {E.}~\bibnamefont {Kroupp}},\ and\ \bibinfo
		{author} {\bibfnamefont {Y.}~\bibnamefont {Maron}},\ }\bibfield  {title}
	{\bibinfo {title} {Spectroscopic determination of the magnetic-field
			distribution in an imploding plasma},\ }\href
	{https://doi.org/10.1063/1.872637} {\bibfield  {journal} {\bibinfo  {journal}
			{Physics of Plasmas}\ }\textbf {\bibinfo {volume} {5}},\ \bibinfo {pages}
		{1068} (\bibinfo {year} {1998})}\BibitemShut {NoStop}%
	\bibitem [{\citenamefont {Gomez}\ \emph
		{et~al.}(2014{\natexlab{b}})\citenamefont {Gomez}, \citenamefont {Hansen},
		\citenamefont {Peterson}, \citenamefont {Bliss}, \citenamefont {Carlson},
		\citenamefont {Lamppa}, \citenamefont {Schroen},\ and\ \citenamefont
		{Rochau}}]{gomez2014zeeman}%
	\BibitemOpen
	\bibfield  {author} {\bibinfo {author} {\bibfnamefont {M.~R.}\ \bibnamefont
			{Gomez}}, \bibinfo {author} {\bibfnamefont {S.~B.}\ \bibnamefont {Hansen}},
		\bibinfo {author} {\bibfnamefont {K.~J.}\ \bibnamefont {Peterson}}, \bibinfo
		{author} {\bibfnamefont {D.~E.}\ \bibnamefont {Bliss}}, \bibinfo {author}
		{\bibfnamefont {A.~L.}\ \bibnamefont {Carlson}}, \bibinfo {author}
		{\bibfnamefont {D.~C.}\ \bibnamefont {Lamppa}}, \bibinfo {author}
		{\bibfnamefont {D.~G.}\ \bibnamefont {Schroen}},\ and\ \bibinfo {author}
		{\bibfnamefont {G.~A.}\ \bibnamefont {Rochau}},\ }\bibfield  {title}
	{\bibinfo {title} {Magnetic field measurements via visible spectroscopy on
			the z machine},\ }\href {https://doi.org/10.1063/1.4891304} {\bibfield
		{journal} {\bibinfo  {journal} {Review of Scientific Instruments}\ }\textbf
		{\bibinfo {volume} {85}},\ \bibinfo {pages} {11E609} (\bibinfo {year}
		{2014}{\natexlab{b}})},\ \Eprint
	{https://arxiv.org/abs/https://doi.org/10.1063/1.4891304}
	{https://doi.org/10.1063/1.4891304} \BibitemShut {NoStop}%
	\bibitem [{\citenamefont {Mikitchuk}\ \emph {et~al.}(2019)\citenamefont
		{Mikitchuk}, \citenamefont {Cveji\ifmmode~\acute{c}\else \'{c}\fi{}},
		\citenamefont {Doron}, \citenamefont {Kroupp}, \citenamefont {Stollberg},
		\citenamefont {Maron}, \citenamefont {Velikovich}, \citenamefont {Ouart},
		\citenamefont {Giuliani}, \citenamefont {Mehlhorn}, \citenamefont {Yu},\ and\
		\citenamefont {Fruchtman}}]{mikitchuk2019}%
	\BibitemOpen
	\bibfield  {author} {\bibinfo {author} {\bibfnamefont {D.}~\bibnamefont
			{Mikitchuk}}, \bibinfo {author} {\bibfnamefont {M.}~\bibnamefont
			{Cveji\ifmmode~\acute{c}\else \'{c}\fi{}}}, \bibinfo {author} {\bibfnamefont
			{R.}~\bibnamefont {Doron}}, \bibinfo {author} {\bibfnamefont
			{E.}~\bibnamefont {Kroupp}}, \bibinfo {author} {\bibfnamefont
			{C.}~\bibnamefont {Stollberg}}, \bibinfo {author} {\bibfnamefont
			{Y.}~\bibnamefont {Maron}}, \bibinfo {author} {\bibfnamefont {A.~L.}\
			\bibnamefont {Velikovich}}, \bibinfo {author} {\bibfnamefont {N.~D.}\
			\bibnamefont {Ouart}}, \bibinfo {author} {\bibfnamefont {J.~L.}\ \bibnamefont
			{Giuliani}}, \bibinfo {author} {\bibfnamefont {T.~A.}\ \bibnamefont
			{Mehlhorn}}, \bibinfo {author} {\bibfnamefont {E.~P.}\ \bibnamefont {Yu}},\
		and\ \bibinfo {author} {\bibfnamefont {A.}~\bibnamefont {Fruchtman}},\
	}\bibfield  {title} {\bibinfo {title} {Effects of a preembedded axial
			magnetic field on the current distribution in a $z$-pinch implosion},\ }\href
	{https://doi.org/10.1103/PhysRevLett.122.045001} {\bibfield  {journal}
		{\bibinfo  {journal} {Phys. Rev. Lett.}\ }\textbf {\bibinfo {volume} {122}},\
		\bibinfo {pages} {045001} (\bibinfo {year} {2019})}\BibitemShut {NoStop}%
	\bibitem [{\citenamefont {Aybar}\ \emph {et~al.}(2022)\citenamefont {Aybar},
		\citenamefont {Conti}, \citenamefont {Cvejic}, \citenamefont {Mikitchuk},
		\citenamefont {Dozieres}, \citenamefont {Kroupp}, \citenamefont {Narkis},
		\citenamefont {Maron},\ and\ \citenamefont {Beg}}]{aybar2022}%
	\BibitemOpen
	\bibfield  {author} {\bibinfo {author} {\bibfnamefont {N.~A.}\ \bibnamefont
			{Aybar}}, \bibinfo {author} {\bibfnamefont {F.}~\bibnamefont {Conti}},
		\bibinfo {author} {\bibfnamefont {M.}~\bibnamefont {Cvejic}}, \bibinfo
		{author} {\bibfnamefont {D.}~\bibnamefont {Mikitchuk}}, \bibinfo {author}
		{\bibfnamefont {M.}~\bibnamefont {Dozieres}}, \bibinfo {author}
		{\bibfnamefont {E.}~\bibnamefont {Kroupp}}, \bibinfo {author} {\bibfnamefont
			{J.}~\bibnamefont {Narkis}}, \bibinfo {author} {\bibfnamefont
			{Y.}~\bibnamefont {Maron}},\ and\ \bibinfo {author} {\bibfnamefont {F.~N.}\
			\bibnamefont {Beg}},\ }\bibfield  {title} {\bibinfo {title} {Dependence of
			plasma-current coupling on current rise time in gas-puff z-pinches},\ }\href
	{https://doi.org/10.1109/TPS.2022.3198385} {\bibfield  {journal} {\bibinfo
			{journal} {IEEE Transactions on Plasma Science}\ }\textbf {\bibinfo {volume}
			{50}},\ \bibinfo {pages} {2541} (\bibinfo {year} {2022})}\BibitemShut
	{NoStop}%
	\bibitem [{\citenamefont {Rosenzweig}\ \emph {et~al.}(2017)\citenamefont
		{Rosenzweig}, \citenamefont {Kroupp}, \citenamefont {Fisher},\ and\
		\citenamefont {Maron}}]{rosenzweig2017jinst}%
	\BibitemOpen
	\bibfield  {author} {\bibinfo {author} {\bibfnamefont {G.}~\bibnamefont
			{Rosenzweig}}, \bibinfo {author} {\bibfnamefont {E.}~\bibnamefont {Kroupp}},
		\bibinfo {author} {\bibfnamefont {A.}~\bibnamefont {Fisher}},\ and\ \bibinfo
		{author} {\bibfnamefont {Y.}~\bibnamefont {Maron}},\ }\bibfield  {title}
	{\bibinfo {title} {Measurements of the spatial magnetic field distribution in
			a z-pinch plasma throughout the stagnation process},\ }\href
	{https://doi.org/10.1088/1748-0221/12/09/P09004} {\bibfield  {journal}
		{\bibinfo  {journal} {Journal of Instrumentation}\ }\textbf {\bibinfo
			{volume} {12}}\bibinfo  {number} { (2017)}}\BibitemShut {NoStop}%
	\bibitem [{\citenamefont {Rosenzweig}\ \emph {et~al.}(2020)\citenamefont
		{Rosenzweig}, \citenamefont {Kroupp}, \citenamefont {Queller}, \citenamefont
		{Starobinets}, \citenamefont {Maron}, \citenamefont {Tangri}, \citenamefont
		{Giuliani},\ and\ \citenamefont {Fruchtman}}]{rosenzweig2020pop}%
	\BibitemOpen
	\bibfield  {number} {  }\bibfield  {author} {\bibinfo {author} {\bibfnamefont
			{G.}~\bibnamefont {Rosenzweig}}, \bibinfo {author} {\bibfnamefont
			{E.}~\bibnamefont {Kroupp}}, \bibinfo {author} {\bibfnamefont
			{T.}~\bibnamefont {Queller}}, \bibinfo {author} {\bibfnamefont
			{A.}~\bibnamefont {Starobinets}}, \bibinfo {author} {\bibfnamefont
			{Y.}~\bibnamefont {Maron}}, \bibinfo {author} {\bibfnamefont
			{V.}~\bibnamefont {Tangri}}, \bibinfo {author} {\bibfnamefont {J.~L.}\
			\bibnamefont {Giuliani}},\ and\ \bibinfo {author} {\bibfnamefont
			{A.}~\bibnamefont {Fruchtman}},\ }\bibfield  {title} {\bibinfo {title} {Local
			measurements of the spatial magnetic field distribution in a z-pinch plasma
			during and near stagnation using polarization spectroscopy},\ }\href
	{https://doi.org/10.1063/1.5126934} {\bibfield  {journal} {\bibinfo
			{journal} {Physics of Plasmas}\ }\textbf {\bibinfo {volume} {27}},\ \bibinfo
		{pages} {022705} (\bibinfo {year} {2020})},\ \Eprint
	{https://arxiv.org/abs/https://doi.org/10.1063/1.5126934}
	{https://doi.org/10.1063/1.5126934} \BibitemShut {NoStop}%
	\bibitem [{\citenamefont {Ivanov}\ \emph {et~al.}(2015)\citenamefont {Ivanov},
		\citenamefont {Anderson}, \citenamefont {Papp}, \citenamefont
		{Astanovitskiy}, \citenamefont {Nalajala},\ and\ \citenamefont
		{Dmitriev}}]{ivanov2015faraday}%
	\BibitemOpen
	\bibfield  {author} {\bibinfo {author} {\bibfnamefont {V.~V.}\ \bibnamefont
			{Ivanov}}, \bibinfo {author} {\bibfnamefont {A.~A.}\ \bibnamefont
			{Anderson}}, \bibinfo {author} {\bibfnamefont {D.}~\bibnamefont {Papp}},
		\bibinfo {author} {\bibfnamefont {A.~L.}\ \bibnamefont {Astanovitskiy}},
		\bibinfo {author} {\bibfnamefont {V.}~\bibnamefont {Nalajala}},\ and\
		\bibinfo {author} {\bibfnamefont {O.}~\bibnamefont {Dmitriev}},\ }\bibfield
	{title} {\bibinfo {title} {Study of magnetic fields and current in the z
			pinch at stagnation},\ }\href {https://doi.org/10.1063/1.4931079} {\bibfield
		{journal} {\bibinfo  {journal} {Physics of Plasmas}\ }\textbf {\bibinfo
			{volume} {22}},\ \bibinfo {pages} {092710} (\bibinfo {year} {2015})},\
	\Eprint {https://arxiv.org/abs/https://doi.org/10.1063/1.4931079}
	{https://doi.org/10.1063/1.4931079} \BibitemShut {NoStop}%
	\bibitem [{\citenamefont {Munzar}\ \emph {et~al.}(2021)\citenamefont {Munzar},
		\citenamefont {Klir}, \citenamefont {Cikhardt}, \citenamefont {Kravarik},
		\citenamefont {Kubes}, \citenamefont {Malir}, \citenamefont {Novotny},
		\citenamefont {Rezac}, \citenamefont {Shishlov}, \citenamefont {Kokshenev},
		\citenamefont {Cherdizov},\ and\ \citenamefont
		{Ratakhin}}]{munzar2021deflectometry}%
	\BibitemOpen
	\bibfield  {author} {\bibinfo {author} {\bibfnamefont {V.}~\bibnamefont
			{Munzar}}, \bibinfo {author} {\bibfnamefont {D.}~\bibnamefont {Klir}},
		\bibinfo {author} {\bibfnamefont {J.}~\bibnamefont {Cikhardt}}, \bibinfo
		{author} {\bibfnamefont {J.}~\bibnamefont {Kravarik}}, \bibinfo {author}
		{\bibfnamefont {P.}~\bibnamefont {Kubes}}, \bibinfo {author} {\bibfnamefont
			{J.}~\bibnamefont {Malir}}, \bibinfo {author} {\bibfnamefont
			{J.}~\bibnamefont {Novotny}}, \bibinfo {author} {\bibfnamefont
			{K.}~\bibnamefont {Rezac}}, \bibinfo {author} {\bibfnamefont {A.~V.}\
			\bibnamefont {Shishlov}}, \bibinfo {author} {\bibfnamefont {V.~A.}\
			\bibnamefont {Kokshenev}}, \bibinfo {author} {\bibfnamefont {R.~K.}\
			\bibnamefont {Cherdizov}},\ and\ \bibinfo {author} {\bibfnamefont {N.~A.}\
			\bibnamefont {Ratakhin}},\ }\bibfield  {title} {\bibinfo {title} {Mapping of
			azimuthal b-fields in z-pinch plasmas using z-pinch-driven ion
			deflectometry},\ }\href {https://doi.org/10.1063/5.0040515} {\bibfield
		{journal} {\bibinfo  {journal} {Physics of Plasmas}\ }\textbf {\bibinfo
			{volume} {28}},\ \bibinfo {pages} {062702} (\bibinfo {year} {2021})},\
	\Eprint {https://arxiv.org/abs/https://doi.org/10.1063/5.0040515}
	{https://doi.org/10.1063/5.0040515} \BibitemShut {NoStop}%
	\bibitem [{\citenamefont {Chang}\ \emph {et~al.}(1991)\citenamefont {Chang},
		\citenamefont {Fisher},\ and\ \citenamefont
		{Van~Drie}}]{chang1991gasDistribution}%
	\BibitemOpen
	\bibfield  {author} {\bibinfo {author} {\bibfnamefont {T.}~\bibnamefont
			{Chang}}, \bibinfo {author} {\bibfnamefont {A.}~\bibnamefont {Fisher}},\ and\
		\bibinfo {author} {\bibfnamefont {A.}~\bibnamefont {Van~Drie}},\ }\bibfield
	{title} {\bibinfo {title} {X-ray results from a modified nozzle and double
			gas puff z pinch},\ }\href {https://doi.org/10.1063/1.348528} {\bibfield
		{journal} {\bibinfo  {journal} {Journal of Applied Physics}\ }\textbf
		{\bibinfo {volume} {69}},\ \bibinfo {pages} {3447} (\bibinfo {year}
		{1991})},\ \Eprint {https://arxiv.org/abs/https://doi.org/10.1063/1.348528}
	{https://doi.org/10.1063/1.348528} \BibitemShut {NoStop}%
	\bibitem [{\citenamefont {Velikovich}\ \emph {et~al.}(1998)\citenamefont
		{Velikovich}, \citenamefont {Cochran}, \citenamefont {Davis},\ and\
		\citenamefont {Chong}}]{velikovich1998tailored}%
	\BibitemOpen
	\bibfield  {author} {\bibinfo {author} {\bibfnamefont {A.~L.}\ \bibnamefont
			{Velikovich}}, \bibinfo {author} {\bibfnamefont {F.~L.}\ \bibnamefont
			{Cochran}}, \bibinfo {author} {\bibfnamefont {J.}~\bibnamefont {Davis}},\
		and\ \bibinfo {author} {\bibfnamefont {Y.~K.}\ \bibnamefont {Chong}},\
	}\bibfield  {title} {\bibinfo {title} {Stabilized radiative z-pinch loads
			with tailored density profiles},\ }\href {https://doi.org/10.1063/1.873063}
	{\bibfield  {journal} {\bibinfo  {journal} {Physics of Plasmas}\ }\textbf
		{\bibinfo {volume} {5}},\ \bibinfo {pages} {3377} (\bibinfo {year} {1998})},\
	\Eprint {https://arxiv.org/abs/https://doi.org/10.1063/1.873063}
	{https://doi.org/10.1063/1.873063} \BibitemShut {NoStop}%
	\bibitem [{\citenamefont {Golingo}\ \emph {et~al.}(2010)\citenamefont
		{Golingo}, \citenamefont {Shumlak},\ and\ \citenamefont
		{Hartog}}]{golingo2010NoteZS}%
	\BibitemOpen
	\bibfield  {author} {\bibinfo {author} {\bibfnamefont {R.~P.}\ \bibnamefont
			{Golingo}}, \bibinfo {author} {\bibfnamefont {U.}~\bibnamefont {Shumlak}},\
		and\ \bibinfo {author} {\bibfnamefont {D.~J.~D.}\ \bibnamefont {Hartog}},\
	}\bibfield  {title} {\bibinfo {title} {Note: Zeeman splitting measurements in
			a high-temperature plasma.},\ }\href@noop {} {\bibfield  {journal} {\bibinfo
			{journal} {The Review of scientific instruments}\ }\textbf {\bibinfo {volume}
			{81}},\ \bibinfo {pages} {126104} (\bibinfo {year} {2010})}\BibitemShut
	{NoStop}%
	\bibitem [{\citenamefont {Lebedev}\ \emph {et~al.}(2002)\citenamefont
		{Lebedev}, \citenamefont {Beg}, \citenamefont {Bland}, \citenamefont
		{Chittenden}, \citenamefont {Dangor},\ and\ \citenamefont
		{Haines}}]{lebedev2002snowplow}%
	\BibitemOpen
	\bibfield  {author} {\bibinfo {author} {\bibfnamefont {S.~V.}\ \bibnamefont
			{Lebedev}}, \bibinfo {author} {\bibfnamefont {F.~N.}\ \bibnamefont {Beg}},
		\bibinfo {author} {\bibfnamefont {S.~N.}\ \bibnamefont {Bland}}, \bibinfo
		{author} {\bibfnamefont {J.~P.}\ \bibnamefont {Chittenden}}, \bibinfo
		{author} {\bibfnamefont {A.~E.}\ \bibnamefont {Dangor}},\ and\ \bibinfo
		{author} {\bibfnamefont {M.~G.}\ \bibnamefont {Haines}},\ }\bibfield  {title}
	{\bibinfo {title} {Snowplow-like behavior in the implosion phase of wire
			array z pinches},\ }\href {https://doi.org/10.1063/1.1466466} {\bibfield
		{journal} {\bibinfo  {journal} {Physics of Plasmas}\ }\textbf {\bibinfo
			{volume} {9}},\ \bibinfo {pages} {2293} (\bibinfo {year} {2002})},\ \Eprint
	{https://arxiv.org/abs/https://doi.org/10.1063/1.1466466}
	{https://doi.org/10.1063/1.1466466} \BibitemShut {NoStop}%
	\bibitem [{\citenamefont {Lebedev}\ \emph {et~al.}(2005)\citenamefont
		{Lebedev}, \citenamefont {Ampleford}, \citenamefont {Bland}, \citenamefont
		{Bott}, \citenamefont {Chittenden}, \citenamefont {Goyer}, \citenamefont
		{Jennings}, \citenamefont {Haines}, \citenamefont {Hall}, \citenamefont
		{Hammer}, \citenamefont {Palmer}, \citenamefont {Pikuz}, \citenamefont
		{Shelkovenko},\ and\ \citenamefont {Christoudias}}]{lebedev2005}%
	\BibitemOpen
	\bibfield  {author} {\bibinfo {author} {\bibfnamefont {S.~V.}\ \bibnamefont
			{Lebedev}}, \bibinfo {author} {\bibfnamefont {D.~J.}\ \bibnamefont
			{Ampleford}}, \bibinfo {author} {\bibfnamefont {S.~N.}\ \bibnamefont
			{Bland}}, \bibinfo {author} {\bibfnamefont {S.~C.}\ \bibnamefont {Bott}},
		\bibinfo {author} {\bibfnamefont {J.~P.}\ \bibnamefont {Chittenden}},
		\bibinfo {author} {\bibfnamefont {J.}~\bibnamefont {Goyer}}, \bibinfo
		{author} {\bibfnamefont {C.}~\bibnamefont {Jennings}}, \bibinfo {author}
		{\bibfnamefont {M.~G.}\ \bibnamefont {Haines}}, \bibinfo {author}
		{\bibfnamefont {G.~N.}\ \bibnamefont {Hall}}, \bibinfo {author}
		{\bibfnamefont {D.~A.}\ \bibnamefont {Hammer}}, \bibinfo {author}
		{\bibfnamefont {J.~B.~A.}\ \bibnamefont {Palmer}}, \bibinfo {author}
		{\bibfnamefont {S.~A.}\ \bibnamefont {Pikuz}}, \bibinfo {author}
		{\bibfnamefont {T.~A.}\ \bibnamefont {Shelkovenko}},\ and\ \bibinfo {author}
		{\bibfnamefont {T.}~\bibnamefont {Christoudias}},\ }\bibfield  {title}
	{\bibinfo {title} {Physics of wire array z-pinch implosions: experiments at
			imperial college},\ }\href {https://doi.org/10.1088/0741-3335/47/5a/009}
	{\bibfield  {journal} {\bibinfo  {journal} {Plasma Physics and Controlled
				Fusion}\ }\textbf {\bibinfo {volume} {47}},\ \bibinfo {pages} {A91} (\bibinfo
		{year} {2005})}\BibitemShut {NoStop}%
	\bibitem [{\citenamefont {Waisman}\ \emph {et~al.}(2004)\citenamefont
		{Waisman}, \citenamefont {Cuneo}, \citenamefont {Stygar}, \citenamefont
		{Lemke}, \citenamefont {Struve},\ and\ \citenamefont
		{Wagoner}}]{waisman2004inductance}%
	\BibitemOpen
	\bibfield  {author} {\bibinfo {author} {\bibfnamefont {E.~M.}\ \bibnamefont
			{Waisman}}, \bibinfo {author} {\bibfnamefont {M.~E.}\ \bibnamefont {Cuneo}},
		\bibinfo {author} {\bibfnamefont {W.~A.}\ \bibnamefont {Stygar}}, \bibinfo
		{author} {\bibfnamefont {R.~W.}\ \bibnamefont {Lemke}}, \bibinfo {author}
		{\bibfnamefont {K.~W.}\ \bibnamefont {Struve}},\ and\ \bibinfo {author}
		{\bibfnamefont {T.~C.}\ \bibnamefont {Wagoner}},\ }\bibfield  {title}
	{\bibinfo {title} {Wire array implosion characteristics from determination of
			load inductance on the z pulsed-power accelerator},\ }\href
	{https://doi.org/10.1063/1.1689969} {\bibfield  {journal} {\bibinfo
			{journal} {Physics of Plasmas}\ }\textbf {\bibinfo {volume} {11}},\ \bibinfo
		{pages} {2009} (\bibinfo {year} {2004})},\ \Eprint
	{https://arxiv.org/abs/https://doi.org/10.1063/1.1689969}
	{https://doi.org/10.1063/1.1689969} \BibitemShut {NoStop}%
	\bibitem [{\citenamefont {Cuneo}\ \emph {et~al.}(2005)\citenamefont {Cuneo},
		\citenamefont {Waisman}, \citenamefont {Lebedev}, \citenamefont {Chittenden},
		\citenamefont {Stygar}, \citenamefont {Chandler}, \citenamefont {Vesey},
		\citenamefont {Yu}, \citenamefont {Nash}, \citenamefont {Bliss},
		\citenamefont {Sarkisov}, \citenamefont {Wagoner}, \citenamefont {Bennett},
		\citenamefont {Sinars}, \citenamefont {Porter}, \citenamefont {Simpson},
		\citenamefont {Ruggles}, \citenamefont {Wenger}, \citenamefont {Garasi},
		\citenamefont {Oliver}, \citenamefont {Aragon}, \citenamefont {Fowler},
		\citenamefont {Hettrick}, \citenamefont {Idzorek}, \citenamefont {Johnson},
		\citenamefont {Keller}, \citenamefont {Lazier}, \citenamefont {McGurn},
		\citenamefont {Mehlhorn}, \citenamefont {Moore}, \citenamefont {Nielsen},
		\citenamefont {Pyle}, \citenamefont {Speas}, \citenamefont {Struve},\ and\
		\citenamefont {Torres}}]{cuneo2005}%
	\BibitemOpen
	\bibfield  {author} {\bibinfo {author} {\bibfnamefont {M.~E.}\ \bibnamefont
			{Cuneo}}, \bibinfo {author} {\bibfnamefont {E.~M.}\ \bibnamefont {Waisman}},
		\bibinfo {author} {\bibfnamefont {S.~V.}\ \bibnamefont {Lebedev}}, \bibinfo
		{author} {\bibfnamefont {J.~P.}\ \bibnamefont {Chittenden}}, \bibinfo
		{author} {\bibfnamefont {W.~A.}\ \bibnamefont {Stygar}}, \bibinfo {author}
		{\bibfnamefont {G.~A.}\ \bibnamefont {Chandler}}, \bibinfo {author}
		{\bibfnamefont {R.~A.}\ \bibnamefont {Vesey}}, \bibinfo {author}
		{\bibfnamefont {E.~P.}\ \bibnamefont {Yu}}, \bibinfo {author} {\bibfnamefont
			{T.~J.}\ \bibnamefont {Nash}}, \bibinfo {author} {\bibfnamefont {D.~E.}\
			\bibnamefont {Bliss}}, \bibinfo {author} {\bibfnamefont {G.~S.}\ \bibnamefont
			{Sarkisov}}, \bibinfo {author} {\bibfnamefont {T.~C.}\ \bibnamefont
			{Wagoner}}, \bibinfo {author} {\bibfnamefont {G.~R.}\ \bibnamefont
			{Bennett}}, \bibinfo {author} {\bibfnamefont {D.~B.}\ \bibnamefont {Sinars}},
		\bibinfo {author} {\bibfnamefont {J.~L.}\ \bibnamefont {Porter}}, \bibinfo
		{author} {\bibfnamefont {W.~W.}\ \bibnamefont {Simpson}}, \bibinfo {author}
		{\bibfnamefont {L.~E.}\ \bibnamefont {Ruggles}}, \bibinfo {author}
		{\bibfnamefont {D.~F.}\ \bibnamefont {Wenger}}, \bibinfo {author}
		{\bibfnamefont {C.~J.}\ \bibnamefont {Garasi}}, \bibinfo {author}
		{\bibfnamefont {B.~V.}\ \bibnamefont {Oliver}}, \bibinfo {author}
		{\bibfnamefont {R.~A.}\ \bibnamefont {Aragon}}, \bibinfo {author}
		{\bibfnamefont {W.~E.}\ \bibnamefont {Fowler}}, \bibinfo {author}
		{\bibfnamefont {M.~C.}\ \bibnamefont {Hettrick}}, \bibinfo {author}
		{\bibfnamefont {G.~C.}\ \bibnamefont {Idzorek}}, \bibinfo {author}
		{\bibfnamefont {D.}~\bibnamefont {Johnson}}, \bibinfo {author} {\bibfnamefont
			{K.}~\bibnamefont {Keller}}, \bibinfo {author} {\bibfnamefont {S.~E.}\
			\bibnamefont {Lazier}}, \bibinfo {author} {\bibfnamefont {J.~S.}\
			\bibnamefont {McGurn}}, \bibinfo {author} {\bibfnamefont {T.~A.}\
			\bibnamefont {Mehlhorn}}, \bibinfo {author} {\bibfnamefont {T.}~\bibnamefont
			{Moore}}, \bibinfo {author} {\bibfnamefont {D.~S.}\ \bibnamefont {Nielsen}},
		\bibinfo {author} {\bibfnamefont {J.}~\bibnamefont {Pyle}}, \bibinfo {author}
		{\bibfnamefont {S.}~\bibnamefont {Speas}}, \bibinfo {author} {\bibfnamefont
			{K.~W.}\ \bibnamefont {Struve}},\ and\ \bibinfo {author} {\bibfnamefont
			{J.~A.}\ \bibnamefont {Torres}},\ }\bibfield  {title} {\bibinfo {title}
		{Characteristics and scaling of tungsten-wire-array z-pinch implosion
			dynamics at 20 ma},\ }\href {https://doi.org/10.1103/PhysRevE.71.046406}
	{\bibfield  {journal} {\bibinfo  {journal} {Phys. Rev. E}\ }\textbf {\bibinfo
			{volume} {71}},\ \bibinfo {pages} {046406} (\bibinfo {year}
		{2005})}\BibitemShut {NoStop}%
	\bibitem [{\citenamefont {Hall}\ \emph {et~al.}(2006)\citenamefont {Hall},
		\citenamefont {Pikuz}, \citenamefont {Shelkovenko}, \citenamefont {Bland},
		\citenamefont {Lebedev}, \citenamefont {Ampleford}, \citenamefont {Palmer},
		\citenamefont {Bott}, \citenamefont {Rapley}, \citenamefont {Chittenden},\
		and\ \citenamefont {Apruzese}}]{hall2006}%
	\BibitemOpen
	\bibfield  {author} {\bibinfo {author} {\bibfnamefont {G.~N.}\ \bibnamefont
			{Hall}}, \bibinfo {author} {\bibfnamefont {S.~A.}\ \bibnamefont {Pikuz}},
		\bibinfo {author} {\bibfnamefont {T.~A.}\ \bibnamefont {Shelkovenko}},
		\bibinfo {author} {\bibfnamefont {S.~N.}\ \bibnamefont {Bland}}, \bibinfo
		{author} {\bibfnamefont {S.~V.}\ \bibnamefont {Lebedev}}, \bibinfo {author}
		{\bibfnamefont {D.~J.}\ \bibnamefont {Ampleford}}, \bibinfo {author}
		{\bibfnamefont {J.~B.~A.}\ \bibnamefont {Palmer}}, \bibinfo {author}
		{\bibfnamefont {S.~C.}\ \bibnamefont {Bott}}, \bibinfo {author}
		{\bibfnamefont {J.}~\bibnamefont {Rapley}}, \bibinfo {author} {\bibfnamefont
			{J.~P.}\ \bibnamefont {Chittenden}},\ and\ \bibinfo {author} {\bibfnamefont
			{J.~P.}\ \bibnamefont {Apruzese}},\ }\bibfield  {title} {\bibinfo {title}
		{Structure of stagnated plasma in aluminum wire array z pinches},\ }\href
	{https://doi.org/10.1063/1.2234284} {\bibfield  {journal} {\bibinfo
			{journal} {Physics of Plasmas}\ }\textbf {\bibinfo {volume} {13}},\ \bibinfo
		{pages} {082701} (\bibinfo {year} {2006})},\ \Eprint
	{https://arxiv.org/abs/https://doi.org/10.1063/1.2234284}
	{https://doi.org/10.1063/1.2234284} \BibitemShut {NoStop}%
	\bibitem [{\citenamefont {Burdiak}\ \emph {et~al.}(2013)\citenamefont
		{Burdiak}, \citenamefont {Lebedev}, \citenamefont {Hall}, \citenamefont
		{Harvey-Thompson}, \citenamefont {Suzuki-Vidal}, \citenamefont {Swadling},
		\citenamefont {Khoory}, \citenamefont {Pickworth}, \citenamefont {Bland},
		\citenamefont {de~Grouchy},\ and\ \citenamefont
		{Skidmore}}]{burdiak2013current}%
	\BibitemOpen
	\bibfield  {author} {\bibinfo {author} {\bibfnamefont {G.~C.}\ \bibnamefont
			{Burdiak}}, \bibinfo {author} {\bibfnamefont {S.~V.}\ \bibnamefont
			{Lebedev}}, \bibinfo {author} {\bibfnamefont {G.~N.}\ \bibnamefont {Hall}},
		\bibinfo {author} {\bibfnamefont {A.~J.}\ \bibnamefont {Harvey-Thompson}},
		\bibinfo {author} {\bibfnamefont {F.}~\bibnamefont {Suzuki-Vidal}}, \bibinfo
		{author} {\bibfnamefont {G.~F.}\ \bibnamefont {Swadling}}, \bibinfo {author}
		{\bibfnamefont {E.}~\bibnamefont {Khoory}}, \bibinfo {author} {\bibfnamefont
			{L.}~\bibnamefont {Pickworth}}, \bibinfo {author} {\bibfnamefont {S.~N.}\
			\bibnamefont {Bland}}, \bibinfo {author} {\bibfnamefont {P.}~\bibnamefont
			{de~Grouchy}},\ and\ \bibinfo {author} {\bibfnamefont {J.}~\bibnamefont
			{Skidmore}},\ }\bibfield  {title} {\bibinfo {title} {Determination of the
			inductance of imploding wire array z-pinches using measurements of load
			voltage},\ }\href {https://doi.org/10.1063/1.4794957} {\bibfield  {journal}
		{\bibinfo  {journal} {Physics of Plasmas}\ }\textbf {\bibinfo {volume}
			{20}},\ \bibinfo {pages} {032705} (\bibinfo {year} {2013})},\ \Eprint
	{https://arxiv.org/abs/https://doi.org/10.1063/1.4794957}
	{https://doi.org/10.1063/1.4794957} \BibitemShut {NoStop}%
	\bibitem [{\citenamefont {Stollberg}(2019)}]{stollberg2019phd}%
	\BibitemOpen
	\bibfield  {author} {\bibinfo {author} {\bibfnamefont {C.}~\bibnamefont
			{Stollberg}},\ }\emph {\bibinfo {title} {PhD thesis: Investigation of a
			small-scale self-compressing plasma column, Weizmann Institute of Science}},\
	\href@noop {} {Ph.D. thesis} (\bibinfo {year} {2019})\BibitemShut {NoStop}%
	\bibitem [{\citenamefont {Gregorian}\ \emph
		{et~al.}(2005{\natexlab{a}})\citenamefont {Gregorian}, \citenamefont
		{Kroupp}, \citenamefont {Davara}, \citenamefont {Starobinets}, \citenamefont
		{Fisher}, \citenamefont {Bernshtam}, \citenamefont {Ralchenko}, \citenamefont
		{Maron}, \citenamefont {Fisher},\ and\ \citenamefont
		{Hoffmann}}]{gregorian2005temperature}%
	\BibitemOpen
	\bibfield  {author} {\bibinfo {author} {\bibfnamefont {L.}~\bibnamefont
			{Gregorian}}, \bibinfo {author} {\bibfnamefont {E.}~\bibnamefont {Kroupp}},
		\bibinfo {author} {\bibfnamefont {G.}~\bibnamefont {Davara}}, \bibinfo
		{author} {\bibfnamefont {A.}~\bibnamefont {Starobinets}}, \bibinfo {author}
		{\bibfnamefont {V.}~\bibnamefont {Fisher}}, \bibinfo {author} {\bibfnamefont
			{V.}~\bibnamefont {Bernshtam}}, \bibinfo {author} {\bibfnamefont
			{Y.}~\bibnamefont {Ralchenko}}, \bibinfo {author} {\bibfnamefont
			{Y.}~\bibnamefont {Maron}}, \bibinfo {author} {\bibfnamefont
			{A.}~\bibnamefont {Fisher}},\ and\ \bibinfo {author} {\bibfnamefont
			{D.}~\bibnamefont {Hoffmann}},\ }\bibfield  {title} {\bibinfo {title}
		{Electron-temperature and energy-flow history in an imploding plasma},\
	}\href@noop {} {\bibfield  {journal} {\bibinfo  {journal} {Physical Review
				E}\ }\textbf {\bibinfo {volume} {71}},\ \bibinfo {pages} {056402} (\bibinfo
		{year} {2005}{\natexlab{a}})}\BibitemShut {NoStop}%
	\bibitem [{\citenamefont {Gregorian}\ \emph
		{et~al.}(2005{\natexlab{b}})\citenamefont {Gregorian}, \citenamefont
		{Kroupp}, \citenamefont {Davara}, \citenamefont {Fisher}, \citenamefont
		{Starobinets}, \citenamefont {Bernshtam}, \citenamefont {Fisher},\ and\
		\citenamefont {Maron}}]{gregorian2005ionization}%
	\BibitemOpen
	\bibfield  {author} {\bibinfo {author} {\bibfnamefont {L.}~\bibnamefont
			{Gregorian}}, \bibinfo {author} {\bibfnamefont {E.}~\bibnamefont {Kroupp}},
		\bibinfo {author} {\bibfnamefont {G.}~\bibnamefont {Davara}}, \bibinfo
		{author} {\bibfnamefont {V.~I.}\ \bibnamefont {Fisher}}, \bibinfo {author}
		{\bibfnamefont {A.}~\bibnamefont {Starobinets}}, \bibinfo {author}
		{\bibfnamefont {V.~A.}\ \bibnamefont {Bernshtam}}, \bibinfo {author}
		{\bibfnamefont {A.}~\bibnamefont {Fisher}},\ and\ \bibinfo {author}
		{\bibfnamefont {Y.}~\bibnamefont {Maron}},\ }\bibfield  {title} {\bibinfo
		{title} {Electron density and ionization dynamics in an imploding z-pinch
			plasma},\ }\href {https://doi.org/10.1063/1.2039943} {\bibfield  {journal}
		{\bibinfo  {journal} {Physics of Plasmas}\ }\textbf {\bibinfo {volume}
			{12}},\ \bibinfo {pages} {092704} (\bibinfo {year} {2005}{\natexlab{b}})},\
	\Eprint {https://arxiv.org/abs/https://doi.org/10.1063/1.2039943}
	{https://doi.org/10.1063/1.2039943} \BibitemShut {NoStop}%
	\bibitem [{\citenamefont {Ralchenko}\ and\ \citenamefont
		{Maron}(2001)}]{ralchenko2001nomad}%
	\BibitemOpen
	\bibfield  {author} {\bibinfo {author} {\bibfnamefont {Y.~V.}\ \bibnamefont
			{Ralchenko}}\ and\ \bibinfo {author} {\bibfnamefont {Y.}~\bibnamefont
			{Maron}},\ }\bibfield  {title} {\bibinfo {title} {Accelerated recombination
			due to resonant deexcitation of metastable states},\ }\href
	{https://doi.org/10.1016/S0022-4073(01)00102-9} {\bibfield  {journal}
		{\bibinfo  {journal} {J. Quant. Spectr. Rad. Transfer}\ }\textbf {\bibinfo
			{volume} {71}},\ \bibinfo {pages} {609} (\bibinfo {year} {2001})}\BibitemShut
	{NoStop}%
	\bibitem [{\citenamefont {Stambulchik}\ and\ \citenamefont
		{Maron}(2006)}]{stambulchik2006broadening}%
	\BibitemOpen
	\bibfield  {author} {\bibinfo {author} {\bibfnamefont {E.}~\bibnamefont
			{Stambulchik}}\ and\ \bibinfo {author} {\bibfnamefont {Y.}~\bibnamefont
			{Maron}},\ }\bibfield  {title} {\bibinfo {title} {A study of ion-dynamics and
			correlation effects for spectral line broadening in plasma: {K}-shell
			lines},\ }\href {https://doi.org/10.1016/j.jqsrt.2005.05.058} {\bibfield
		{journal} {\bibinfo  {journal} {J. Quant. Spectr. Rad. Transfer}\ }\textbf
		{\bibinfo {volume} {99}},\ \bibinfo {pages} {730} (\bibinfo {year}
		{2006})}\BibitemShut {NoStop}%
	\bibitem [{\citenamefont {Stambulchik}\ and\ \citenamefont
		{Maron}(2008)}]{stambulchik2008stark}%
	\BibitemOpen
	\bibfield  {author} {\bibinfo {author} {\bibfnamefont {E.}~\bibnamefont
			{Stambulchik}}\ and\ \bibinfo {author} {\bibfnamefont {Y.}~\bibnamefont
			{Maron}},\ }\bibfield  {title} {\bibinfo {title} {Stark effect of high-{$n$}
			hydrogen-like transitions: quasi-contiguous approximation},\ }\href
	{https://doi.org/10.1088/0953-4075/41/9/095703} {\bibfield  {journal}
		{\bibinfo  {journal} {J. Phys. B: At. Mol. Opt. Phys.}\ }\textbf {\bibinfo
			{volume} {41}},\ \bibinfo {pages} {095703} (\bibinfo {year}
		{2008})}\BibitemShut {NoStop}%
	\bibitem [{\citenamefont {Konjevi\'{c}}\ \emph {et~al.}(2002)\citenamefont
		{Konjevi\'{c}}, \citenamefont {Lesage}, \citenamefont {Fuhr},\ and\
		\citenamefont {Wiese}}]{konjevic2002stark}%
	\BibitemOpen
	\bibfield  {author} {\bibinfo {author} {\bibfnamefont {N.}~\bibnamefont
			{Konjevi\'{c}}}, \bibinfo {author} {\bibfnamefont {A.}~\bibnamefont
			{Lesage}}, \bibinfo {author} {\bibfnamefont {J.~R.}\ \bibnamefont {Fuhr}},\
		and\ \bibinfo {author} {\bibfnamefont {W.~L.}\ \bibnamefont {Wiese}},\
	}\bibfield  {title} {\bibinfo {title} {Experimental stark widths and shifts
			for spectral lines of neutral and ionized atoms (a critical review of
			selected data for the period 1989 through 2000)},\ }\href
	{https://doi.org/10.1063/1.1486456} {\bibfield  {journal} {\bibinfo
			{journal} {Journal of Physical and Chemical Reference Data}\ }\textbf
		{\bibinfo {volume} {31}},\ \bibinfo {pages} {819} (\bibinfo {year}
		{2002})}\BibitemShut {NoStop}%
	\bibitem [{\citenamefont {Ochs}\ \emph {et~al.}(2019)\citenamefont {Ochs},
		\citenamefont {Stollberg}, \citenamefont {Kroupp}, \citenamefont {Maron},
		\citenamefont {Fruchtman}, \citenamefont {Kolmes}, \citenamefont {Mlodik},\
		and\ \citenamefont {Fisch}}]{ochs2019current}%
	\BibitemOpen
	\bibfield  {author} {\bibinfo {author} {\bibfnamefont {I.~E.}\ \bibnamefont
			{Ochs}}, \bibinfo {author} {\bibfnamefont {C.}~\bibnamefont {Stollberg}},
		\bibinfo {author} {\bibfnamefont {E.}~\bibnamefont {Kroupp}}, \bibinfo
		{author} {\bibfnamefont {Y.}~\bibnamefont {Maron}}, \bibinfo {author}
		{\bibfnamefont {A.}~\bibnamefont {Fruchtman}}, \bibinfo {author}
		{\bibfnamefont {E.~J.}\ \bibnamefont {Kolmes}}, \bibinfo {author}
		{\bibfnamefont {M.~E.}\ \bibnamefont {Mlodik}},\ and\ \bibinfo {author}
		{\bibfnamefont {N.~J.}\ \bibnamefont {Fisch}},\ }\bibfield  {title} {\bibinfo
		{title} {Current channel evolution in ideal z pinch for general velocity
			profiles},\ }\href {https://doi.org/10.1063/1.5118668} {\bibfield  {journal}
		{\bibinfo  {journal} {Physics of Plasmas}\ }\textbf {\bibinfo {volume}
			{26}},\ \bibinfo {pages} {122706} (\bibinfo {year} {2019})},\ \Eprint
	{https://arxiv.org/abs/https://doi.org/10.1063/1.5118668}
	{https://doi.org/10.1063/1.5118668} \BibitemShut {NoStop}%
	\bibitem [{\citenamefont {Haines}(1959)}]{haines1959skinEffect}%
	\BibitemOpen
	\bibfield  {author} {\bibinfo {author} {\bibfnamefont {M.~G.}\ \bibnamefont
			{Haines}},\ }\bibfield  {title} {\bibinfo {title} {The inverse skin effect},\
	}\href {https://doi.org/10.1088/0370-1328/74/5/310} {\bibfield  {journal}
		{\bibinfo  {journal} {Proc. Phys. Soc.}\ }\textbf {\bibinfo {volume} {74}},\
		\bibinfo {pages} {576} (\bibinfo {year} {1959})}\BibitemShut {NoStop}%
	\bibitem [{\citenamefont {Lieberman}\ and\ \citenamefont
		{Lichtenberg}(2005)}]{lieberman2005principles}%
	\BibitemOpen
	\bibfield  {author} {\bibinfo {author} {\bibfnamefont {M.~A.}\ \bibnamefont
			{Lieberman}}\ and\ \bibinfo {author} {\bibfnamefont {A.~J.}\ \bibnamefont
			{Lichtenberg}},\ }\href@noop {} {\emph {\bibinfo {title} {Principles of
				plasma discharges and materials processing}}}\ (\bibinfo  {publisher} {John
		Wiley \& Sons},\ \bibinfo {year} {2005})\BibitemShut {NoStop}%
	\bibitem [{\citenamefont {Waisman}\ \emph {et~al.}(2008)\citenamefont
		{Waisman}, \citenamefont {Cuneo}, \citenamefont {Lemke}, \citenamefont
		{Sinars},\ and\ \citenamefont {Stygar}}]{waisman2008resistance}%
	\BibitemOpen
	\bibfield  {author} {\bibinfo {author} {\bibfnamefont {E.~M.}\ \bibnamefont
			{Waisman}}, \bibinfo {author} {\bibfnamefont {M.~E.}\ \bibnamefont {Cuneo}},
		\bibinfo {author} {\bibfnamefont {R.~W.}\ \bibnamefont {Lemke}}, \bibinfo
		{author} {\bibfnamefont {D.~B.}\ \bibnamefont {Sinars}},\ and\ \bibinfo
		{author} {\bibfnamefont {W.~A.}\ \bibnamefont {Stygar}},\ }\bibfield  {title}
	{\bibinfo {title} {Lower bounds for the kinetic energy and resistance of wire
			array z pinches on the z pulsed-power accelerator},\ }\href
	{https://doi.org/10.1063/1.2898724} {\bibfield  {journal} {\bibinfo
			{journal} {Physics of Plasmas}\ }\textbf {\bibinfo {volume} {15}},\ \bibinfo
		{pages} {042702} (\bibinfo {year} {2008})},\ \Eprint
	{https://arxiv.org/abs/https://aip.scitation.org/doi/pdf/10.1063/1.2898724}
	{https://aip.scitation.org/doi/pdf/10.1063/1.2898724} \BibitemShut {NoStop}%
	\bibitem [{\citenamefont {Peterkin}\ \emph {et~al.}(1998)\citenamefont
		{Peterkin}, \citenamefont {Frese},\ and\ \citenamefont
		{Sovinec}}]{peterkin1998mach2}%
	\BibitemOpen
	\bibfield  {author} {\bibinfo {author} {\bibfnamefont {R.~E.}\ \bibnamefont
			{Peterkin}}, \bibinfo {author} {\bibfnamefont {M.~H.}\ \bibnamefont
			{Frese}},\ and\ \bibinfo {author} {\bibfnamefont {C.~R.}\ \bibnamefont
			{Sovinec}},\ }\bibfield  {title} {\bibinfo {title} {Transport of magnetic
			flux in an arbitrary coordinate ale code},\ }\href
	{https://doi.org/https://doi.org/10.1006/jcph.1998.5880} {\bibfield
		{journal} {\bibinfo  {journal} {Journal of Computational Physics}\ }\textbf
		{\bibinfo {volume} {140}},\ \bibinfo {pages} {148 } (\bibinfo {year}
		{1998})}\BibitemShut {NoStop}%
	\bibitem [{\citenamefont {Awe}\ \emph {et~al.}(2013)\citenamefont {Awe},
		\citenamefont {McBride}, \citenamefont {Jennings}, \citenamefont {Lamppa},
		\citenamefont {Martin}, \citenamefont {Rovang}, \citenamefont {Slutz},
		\citenamefont {Cuneo}, \citenamefont {Owen}, \citenamefont {Sinars},
		\citenamefont {Tomlinson}, \citenamefont {Gomez}, \citenamefont {Hansen},
		\citenamefont {Herrmann}, \citenamefont {McKenney}, \citenamefont {Nakhleh},
		\citenamefont {Robertson}, \citenamefont {Rochau}, \citenamefont {Savage},
		\citenamefont {Schroen},\ and\ \citenamefont {Stygar}}]{awe2013instability}%
	\BibitemOpen
	\bibfield  {author} {\bibinfo {author} {\bibfnamefont {T.~J.}\ \bibnamefont
			{Awe}}, \bibinfo {author} {\bibfnamefont {R.~D.}\ \bibnamefont {McBride}},
		\bibinfo {author} {\bibfnamefont {C.~A.}\ \bibnamefont {Jennings}}, \bibinfo
		{author} {\bibfnamefont {D.~C.}\ \bibnamefont {Lamppa}}, \bibinfo {author}
		{\bibfnamefont {M.~R.}\ \bibnamefont {Martin}}, \bibinfo {author}
		{\bibfnamefont {D.~C.}\ \bibnamefont {Rovang}}, \bibinfo {author}
		{\bibfnamefont {S.~A.}\ \bibnamefont {Slutz}}, \bibinfo {author}
		{\bibfnamefont {M.~E.}\ \bibnamefont {Cuneo}}, \bibinfo {author}
		{\bibfnamefont {A.~C.}\ \bibnamefont {Owen}}, \bibinfo {author}
		{\bibfnamefont {D.~B.}\ \bibnamefont {Sinars}}, \bibinfo {author}
		{\bibfnamefont {K.}~\bibnamefont {Tomlinson}}, \bibinfo {author}
		{\bibfnamefont {M.~R.}\ \bibnamefont {Gomez}}, \bibinfo {author}
		{\bibfnamefont {S.~B.}\ \bibnamefont {Hansen}}, \bibinfo {author}
		{\bibfnamefont {M.~C.}\ \bibnamefont {Herrmann}}, \bibinfo {author}
		{\bibfnamefont {J.~L.}\ \bibnamefont {McKenney}}, \bibinfo {author}
		{\bibfnamefont {C.}~\bibnamefont {Nakhleh}}, \bibinfo {author} {\bibfnamefont
			{G.~K.}\ \bibnamefont {Robertson}}, \bibinfo {author} {\bibfnamefont {G.~A.}\
			\bibnamefont {Rochau}}, \bibinfo {author} {\bibfnamefont {M.~E.}\
			\bibnamefont {Savage}}, \bibinfo {author} {\bibfnamefont {D.~G.}\
			\bibnamefont {Schroen}},\ and\ \bibinfo {author} {\bibfnamefont {W.~A.}\
			\bibnamefont {Stygar}},\ }\bibfield  {title} {\bibinfo {title} {Observations
			of modified three-dimensional instability structure for imploding z-pinch
			liners that are premagnetized with an axial field},\ }\href
	{https://doi.org/10.1103/PhysRevLett.111.235005} {\bibfield  {journal}
		{\bibinfo  {journal} {Phys. Rev. Lett.}\ }\textbf {\bibinfo {volume} {111}},\
		\bibinfo {pages} {235005} (\bibinfo {year} {2013})}\BibitemShut {NoStop}%
	\bibitem [{\citenamefont {Giuliani}\ \emph {et~al.}(2014)\citenamefont
		{Giuliani}, \citenamefont {Thornhill}, \citenamefont {Kroupp}, \citenamefont
		{Osin}, \citenamefont {Maron}, \citenamefont {Dasgupta}, \citenamefont
		{Apruzese}, \citenamefont {Velikovich}, \citenamefont {Chong}, \citenamefont
		{Starobinets}, \citenamefont {Fisher}, \citenamefont {Zarnitsky},
		\citenamefont {Bernshtam}, \citenamefont {Fisher}, \citenamefont {Mehlhorn},\
		and\ \citenamefont {Deeney}}]{giulliani2014temperatures}%
	\BibitemOpen
	\bibfield  {author} {\bibinfo {author} {\bibfnamefont {J.~L.}\ \bibnamefont
			{Giuliani}}, \bibinfo {author} {\bibfnamefont {J.~W.}\ \bibnamefont
			{Thornhill}}, \bibinfo {author} {\bibfnamefont {E.}~\bibnamefont {Kroupp}},
		\bibinfo {author} {\bibfnamefont {D.}~\bibnamefont {Osin}}, \bibinfo {author}
		{\bibfnamefont {Y.}~\bibnamefont {Maron}}, \bibinfo {author} {\bibfnamefont
			{A.}~\bibnamefont {Dasgupta}}, \bibinfo {author} {\bibfnamefont {J.~P.}\
			\bibnamefont {Apruzese}}, \bibinfo {author} {\bibfnamefont {A.~L.}\
			\bibnamefont {Velikovich}}, \bibinfo {author} {\bibfnamefont {Y.~K.}\
			\bibnamefont {Chong}}, \bibinfo {author} {\bibfnamefont {A.}~\bibnamefont
			{Starobinets}}, \bibinfo {author} {\bibfnamefont {V.}~\bibnamefont {Fisher}},
		\bibinfo {author} {\bibfnamefont {Y.}~\bibnamefont {Zarnitsky}}, \bibinfo
		{author} {\bibfnamefont {V.}~\bibnamefont {Bernshtam}}, \bibinfo {author}
		{\bibfnamefont {A.}~\bibnamefont {Fisher}}, \bibinfo {author} {\bibfnamefont
			{T.~A.}\ \bibnamefont {Mehlhorn}},\ and\ \bibinfo {author} {\bibfnamefont
			{C.}~\bibnamefont {Deeney}},\ }\bibfield  {title} {\bibinfo {title}
		{Effective versus ion thermal temperatures in the weizmann ne z-pinch:
			Modeling and stagnation physics},\ }\href {https://doi.org/10.1063/1.4865223}
	{\bibfield  {journal} {\bibinfo  {journal} {Physics of Plasmas}\ }\textbf
		{\bibinfo {volume} {21}},\ \bibinfo {pages} {031209} (\bibinfo {year}
		{2014})},\ \Eprint {https://arxiv.org/abs/https://doi.org/10.1063/1.4865223}
	{https://doi.org/10.1063/1.4865223} \BibitemShut {NoStop}%
	\bibitem [{\citenamefont {Mitrofanov}\ \emph {et~al.}(2017)\citenamefont
		{Mitrofanov}, \citenamefont {Aleksandrov}, \citenamefont {Grabovski},
		\citenamefont {Branitsky}, \citenamefont {Gritsuk}, \citenamefont {Frolov},\
		and\ \citenamefont {Laukhin}}]{mitrofanov2017trailing}%
	\BibitemOpen
	\bibfield  {author} {\bibinfo {author} {\bibfnamefont {K.~N.}\ \bibnamefont
			{Mitrofanov}}, \bibinfo {author} {\bibfnamefont {V.~V.}\ \bibnamefont
			{Aleksandrov}}, \bibinfo {author} {\bibfnamefont {E.~V.}\ \bibnamefont
			{Grabovski}}, \bibinfo {author} {\bibfnamefont {A.~V.}\ \bibnamefont
			{Branitsky}}, \bibinfo {author} {\bibfnamefont {A.~N.}\ \bibnamefont
			{Gritsuk}}, \bibinfo {author} {\bibfnamefont {I.~N.}\ \bibnamefont
			{Frolov}},\ and\ \bibinfo {author} {\bibfnamefont {Y.~N.}\ \bibnamefont
			{Laukhin}},\ }\bibfield  {title} {\bibinfo {title} {Stability of compression
			of the inner array plasma in nested arrays},\ }\href
	{https://doi.org/10.1134/S1063780X17090082} {\bibfield  {journal} {\bibinfo
			{journal} {Plasma Phys. Rep.}\ }\textbf {\bibinfo {volume} {43}} (\bibinfo
		{year} {2017})}\BibitemShut {NoStop}%
	\bibitem [{\citenamefont {Rozo}\ \emph {et~al.}(2019)\citenamefont {Rozo},
		\citenamefont {Utz}, \citenamefont {Vargas~Dom\'{\i}nguez}, \citenamefont
		{Veronig},\ and\ \citenamefont
		{Van~Doorsselaere}}]{camposRozo2019solarBfield}%
	\BibitemOpen
	\bibfield  {author} {\bibinfo {author} {\bibfnamefont {J.~I.}\ \bibnamefont
			{Rozo}}, \bibinfo {author} {\bibfnamefont {D.}~\bibnamefont {Utz}}, \bibinfo
		{author} {\bibfnamefont {S.}~\bibnamefont {Vargas~Dom\'{\i}nguez}}, \bibinfo
		{author} {\bibfnamefont {A.}~\bibnamefont {Veronig}},\ and\ \bibinfo {author}
		{\bibfnamefont {T.}~\bibnamefont {Van~Doorsselaere}},\ }\bibfield  {title}
	{\bibinfo {title} {Photospheric plasma and magnetic field dynamics during the
			formation of solar ar 11190},\ }\href
	{https://doi.org/10.1051/0004-6361/201832760} {\bibfield  {journal} {\bibinfo
			{journal} {A\&A}\ }\textbf {\bibinfo {volume} {622}},\ \bibinfo {pages}
		{A168} (\bibinfo {year} {2019})}\BibitemShut {NoStop}%
	\bibitem [{\citenamefont {Jarboe}\ \emph {et~al.}(2019)\citenamefont {Jarboe},
		\citenamefont {Benedett}, \citenamefont {Everson}, \citenamefont {Hansen},
		\citenamefont {Hossack}, \citenamefont {Morgan}, \citenamefont {Nelson},
		\citenamefont {O'Bryan}, \citenamefont {Penna},\ and\ \citenamefont
		{Sutherland}}]{jarboe2019solarBfield}%
	\BibitemOpen
	\bibfield  {author} {\bibinfo {author} {\bibfnamefont {T.~R.}\ \bibnamefont
			{Jarboe}}, \bibinfo {author} {\bibfnamefont {T.~E.}\ \bibnamefont
			{Benedett}}, \bibinfo {author} {\bibfnamefont {C.~J.}\ \bibnamefont
			{Everson}}, \bibinfo {author} {\bibfnamefont {C.~J.}\ \bibnamefont {Hansen}},
		\bibinfo {author} {\bibfnamefont {A.~C.}\ \bibnamefont {Hossack}}, \bibinfo
		{author} {\bibfnamefont {K.~D.}\ \bibnamefont {Morgan}}, \bibinfo {author}
		{\bibfnamefont {B.~A.}\ \bibnamefont {Nelson}}, \bibinfo {author}
		{\bibfnamefont {J.~B.}\ \bibnamefont {O'Bryan}}, \bibinfo {author}
		{\bibfnamefont {J.~M.}\ \bibnamefont {Penna}},\ and\ \bibinfo {author}
		{\bibfnamefont {D.~A.}\ \bibnamefont {Sutherland}},\ }\bibfield  {title}
	{\bibinfo {title} {The nature and source of solar magnetic phenomena},\
	}\href {https://doi.org/10.1063/1.5087613} {\bibfield  {journal} {\bibinfo
			{journal} {Physics of Plasmas}\ }\textbf {\bibinfo {volume} {26}},\ \bibinfo
		{pages} {092902} (\bibinfo {year} {2019})},\ \Eprint
	{https://arxiv.org/abs/https://doi.org/10.1063/1.5087613}
	{https://doi.org/10.1063/1.5087613} \BibitemShut {NoStop}%
\end{thebibliography}

%

\end{document}